**Evidence for unconventional superconductivity in twisted bilayer graphene**

Myungchul Oh[1,*], Kevin P. Nuckolls[1,*], Dillon Wong[1,*], Ryan L. Lee[1], Xiaomeng Liu[1], Kenji Watanabe[2], Takashi Taniguchi[3], Ali Yazdani[1,‡]

[1]Joseph Henry Laboratories and Department of Physics, Princeton University, Princeton, NJ 08544, USA

[2]Research Center for Functional Materials, National Institute for Materials Science, 1-1 Namiki, Tsukuba 305-0044, Japan

[3]International Center for Materials Nanoarchitectonics, National Institute for Materials Science, 1-1 Namiki, Tsukuba 305-0044, Japan

* These authors contributed equally to this work.
‡ Corresponding author email: yazdani@princeton.edu

**The emergence of superconductivity and correlated insulators in magic-angle twisted bilayer graphene (MATBG) has raised the intriguing possibility that its pairing mechanism is distinct from that of conventional superconductors[1–4], as described by the Bardeen-Cooper-Schrieffer (BCS) theory. However, recent studies have shown that superconductivity persists even when Coulomb interactions are partially screened[5,6]. This suggests that pairing in MATBG might be conventional in nature and a consequence of the large density of states (DOS) of its flat bands. Here we combine tunneling and Andreev reflection spectroscopy with the scanning tunneling microscope (STM) to observe several key experimental signatures for unconventional superconductivity in MATBG. We show that the tunneling spectra below the transition temperature $T_c$ are inconsistent with those of a conventional s-wave superconductor, but rather resemble those of a nodal superconductor with an anisotropic pairing mechanism. We observe a large discrepancy between the tunneling gap $\Delta_T$, which far exceeds the mean-field BCS ratio (with $2\Delta_T/k_B T_c \sim 25$), and the gap $\Delta_{AR}$ extracted from Andreev reflection spectroscopy ($2\Delta_{AR}/k_B T_c \sim 6$). The tunneling gap persists even when superconductivity is suppressed, indicating its emergence from a pseudogap phase. Moreover, the pseudogap and superconductivity are both absent when MATBG is aligned with hexagonal boron nitride (hBN). These findings and other observations reported here provide a preponderance of evidence for a non-BCS mechanism for superconductivity in MATBG.**

Tunneling measurements of the quasiparticle DOS, the energy gap, and electron-phonon coupling in conventional superconductors have provided key experimental evidence for the BCS theory of superconductivity[7]. Similar measurements on correlated superconductors, most notably STM and angle-resolved photoemission spectroscopy, have shown their properties to be qualitatively different from those of BCS superconductors[8,9]. For the high-$T_c$ cuprate superconductors, while tunneling spectra in the overdoped regime can be captured by the DOS of a BCS-like model with a d-wave order parameter, the yet-to-be-understood pseudogap phenomenon at reduced doping causes the spectroscopic properties of the cuprates to strongly deviate from this picture[9].

Superconductivity has been observed at remarkably low carrier densities at partial fillings of the flat bands of MATBG[1–3]. Although these qualities suggest an unconventional pairing mechanism, conclusive evidence for any mechanism beyond the BCS paradigm is absent. We use density-tuned scanning tunneling and point-contact spectroscopy (DT-STS, DT-PCS) to show that the superconducting phase of MATBG, specifically when hole-doping its flat valence band, shares a remarkable number of features with unconventional superconductors. Our experiments show a V-shaped gap at low temperatures and an unusual pseudogap precursor phase at higher temperatures and magnetic fields from which phase-coherent superconductivity emerges. The low-energy region of the V-shaped gap supports an anisotropic pairing mechanism with nodes in the superconducting gap function, as anticipated by some theoretical studies[10–12]. The pseudogap state may signify pairing without phase coherence or a secondary phase forming above $T_c$ and $B_c$. Both the pseudogap and superconductivity are absent when MATBG is commensurately aligned with the hBN substrate, suggesting that the structural characteristics and/or the $C_2T$-symmetry of unaligned MATBG are required for stabilizing these ground states. Although we cannot rule out a phonon-based pairing mechanism[13,14], our results provide key constraints for an accurate theory of superconductivity in MATBG.

We performed our experiments in a homebuilt dilution-refrigerator STM[15] on devices sketched in Fig. 1a. MATBG, biased at $V_s$, rests on hBN/SiO$_2$/Si, while a voltage $V_g$ applied to Si tunes the carrier density (see Methods). Fig. 1b shows a topographic image[16–18] of unaligned MATBG/hBN, while Fig. 1c shows DT-STS d$I$/d$V(V_s, V_g)$ acquired at 250 mK at the center of an AA site in Device A (see SI for AB/BA data). Fig. 1c shows that the conduction (valence) flat band is pinned to the Fermi energy (E$_F$; $V_s = 0$ V) when $V_g$ tunes E$_F$ above (below) the charge neutrality point (CNP; $V_{CNP} = 3.7$ V), while the valence (conduction) flat band onsets at $V_s < -20$ mV ($V_s > 20$ mV) and displays significant energy broadening due to charge fluctuations[19]. At millikelvin temperatures, we observe features in DT-STS attributed to a cascade of transitions at partial band fillings[20–23], but they appear weaker and broader in energy than those observed at higher temperatures (T > 4 K), which may be related to high-entropy isospin fluctuations[24,25].

**Distinguishing nodal superconductivity**

DT-STS shows several gapped phases (Fig. 1c) starting with band insulators at $\nu = \pm 4$, where $\nu$ is the electron filling per moiré unit cell relative to the CNP. Here we focus on partial fillings of the valence flat band near $-3 < \nu < -2$, where transport studies[1–3] report superconductivity in MATBG (Fig. 1c red box). Figures 1d and 1f show tunneling spectra $dI/dV(V_s, V_g)$ from two devices (Device A as in Fig. 1c, and Device B), which display a gap at $\nu = -2$ that opens and closes at $E_F$ with decreasing $V_g$, followed by the opening of a new gap that persists between $-3 < \nu < -2$. The density dependence of these gaps is highlighted by $dI/dV(V_s = 0\ V)$ as a function of $V_g$ shown in Figs. 1d and 1f. We observe a clear transition between the two gapped phases, consistent with the phase diagram of MATBG from transport studies[1–3,5,6] in which a correlated insulator (CI) at $\nu = -2$ transitions into a superconductor that persists for $-3 < \nu < -2$. In Figs. 1e and 1g, we plot tunneling gaps for the $\nu = -2$ CI (red curves) and the $-3 < \nu < -2$ superconductor (blue curves) measured in each device. The $-3 < \nu < -2$ tunneling gap is significantly larger than $k_B T_c$ observed in transport experiments[1–3] and is an order of magnitude larger than an in-plane tunneling gap observed in an MATBG p-n junction[26] (presumably, the lateral p-n junction only probes the edge of the superconducting dome adjacent to the CI, instead of optimal doping, due to the junction's doping gradient). Before examining the shapes of the tunneling spectra further, we discuss our method for distinguishing between gapped insulating and superconducting phases by complementing DT-STS with point-contact spectroscopy (PCS) measurements.

Because both CIs and superconductors show suppressions in $dI/dV(V_s = 0\ V)$, we require complementary information that distinguishes these two phases. We performed PCS by reducing the tip height above the sample until the tip makes point contact with the sample surface (sketched in Fig. 2a) and then measuring the two-terminal tip-sample conductance $G(V_s, V_g)$ (see Methods; see SI for discussion of possible tip-induced pressure and strain during PCS). This measurement is particularly sensitive to the local region beneath the tip (see SI). The point-contact zero-bias conductance $G(V_s = 0\ V, V_g)$ (PCS-ZBC), plotted in Fig. 2b (Device A' - a different region of Device A) as a function of $V_g$, vanishes at $\nu = \pm 4$ and $\nu = \pm 2$, signaling the formation of band and correlated insulators (Fig. 2b; red shaded bars). Consistent with transport studies[1–3], the insulating states are insensitive to application of a weak out-of-plane magnetic field B. In contrast, the PCS-ZBC displays enhanced intensity between $-3 < \nu < -2$ (Fig. 2b; green shaded bar) that is suppressed with increasing B, consistent with superconductivity in this doping range. More direct evidence for superconductivity is revealed by the voltage-bias dependence of PCS $G(V_s, V_g)$ in Figs. 2c-d. These spectra are indicative of Andreev reflection[27,28], where incoming electrons from the metallic tip are reflected as holes while Cooper pairs propagate into the superconducting sample (Fig. 2a). This results in enhanced conductance at low biases and "excess current" when the sample is superconducting. Signatures of Andreev reflection in PCS $G(V_s, V_g)$ (black boxes in Figs. 2e-f) are limited roughly to fillings $-3 < \nu < -2$, magnetic fields $B < B_c \sim 50$ mT, and temperatures $T < T_c \sim 1.2$ K, all of which are consistent with transport measurements[1–3]. A side-by-side comparison of STS and PCS (Fig. 2g) at the same sample location shows how PCS can clearly distinguish tunneling gaps associated with superconductivity from those associated with insulators. Despite the presence of many correlation-driven gaps at $E_F$ in STS, only the filling range $-3 < \nu < -2$ shows both a V-shaped gap in STS and a zero-bias conductance peak in PCS.

**Two distinct energy scales & the pseudogap**

Both STS and PCS provide complementary evidence for an anisotropic pairing mechanism of superconductivity in MATBG. Moreover, these measurements establish two distinct energy scales. Low-energy STS spectra (Fig. 3a) are clearly incompatible with an isotropic s-wave pairing symmetry, and the best fits to such a model require introducing unphysically large quasiparticle broadening (equivalent to an electron temperature above 2 K; for comparison, see SI for STS on superconducting Al). Often STS spectra on MATBG have a finite conductance at zero energy, but V-shaped spectra with zero conductance at zero bias have also been observed (Fig. 1g). These STS spectra resemble the quasiparticle DOS of a nodal superconductor, as for higher angular momentum (e.g. p- or d-wave) pairing with an anisotropic gap function (Fig. 3b shows this fit for Device A, $V_g = -25.8$ V – see SI for fits at other $V_g$). Although the nodal fit describes this spectrum well, one should be cautious about this interpretation given the similar appearance of this gap to that of the pseudogap above $T_c$ and $B_c$ described below. Nevertheless, we extract an energy scale of $\Delta_T \sim 0.9$ meV from this fit, which roughly corresponds to half the separation of the shoulders in the spectrum. Similarly, the Andreev reflection spectra in PCS resemble predictions from the Blonder-Tinkham-Klapwijk (BTK) model[29] using a nodal superconducting gap function (Fig. 3c and SI). However, a BTK-model fit yields an energy scale $\Delta_{AR} \sim 0.3$ meV (Device A': $V_g = -22.8$ V), 3 - 5 times smaller than $\Delta_T$. For $T_c \approx 1.2$ K (measured through PCS), the observed ratio $2\Delta_T/k_B T_c \sim 25$ (Device A';

$V_g$ = -22.8 V) is significantly higher than the expected ratio for a tunneling gap of a BCS superconductor ($2\Delta_{MF}/k_BT_c$ = 3.53). The Andreev energy-scale ratio $2\Delta_{AR}/k_BT_c$ ~ 6 also appears to be higher than the BCS ratio. As noted above, Andreev reflection disappears when phase-coherent superconductivity is absent, with both $\Delta_{AR}$ and the Andreev excess current vanishing above $T_c$ and $B_c$ (Figs. 2d and 3d). In contrast, the STS gap $\Delta_T$ persists when phase-coherent superconductivity vanishes above $T_c$ (see Fig. 3e and SI) and well-above $B_c$ (Fig. 4).

A similar dichotomy between the energy scales describing tunneling and Andreev reflection has been documented for the underdoped cuprate superconductors[27], where Andreev reflection also tracks the onset of phase coherence at $T_c$ while the tunneling gap persists above $T_c$, as we observe in MATBG (see high-temperature data in SI; see also Ref. [23]). Compared to studies[8,9] that examine the relationship between the pseudogap and superconductivity in the cuprates, in MATBG, we have the advantage that application of a relatively weak B > $B_c$ ~ 50 mT suppresses phase coherence at low temperatures, allowing us to probe the shape of the pseudogap spectra with high energy resolution at the lowest temperatures. Such measurements in Fig. 4a show that the shapes of the spectra in the pseudogap phase remain remarkably sharp and surprisingly similar to those of spectra observed when the sample is superconducting. While the $\nu$ = -2 CI is suppressed below 3 T, the pseudogap remains present over most of the doping range -3 < $\nu$ < -2 (Fig. 4b). The density dependence of the STS at zero magnetic field and up to 3 T (see SI for 1 T data) reveals that the onset of a sharp pseudogap in the absence of phase-coherent superconductivity occurs when the van Hove singularity associated with the valence band overlaps with $E_F$. In this situation, the gain in the exchange energy may favor the formation of an isospin (spin/valley) polarized/coherent ground state (or some other ordered state), which may be responsible for the pseudogap with sharp side-peaks shown in Fig. 4a. However, given the remarkable resemblance between the shapes of the STS gaps in the pseudogap and superconducting phases, it is also possible that such a gap is driven by the formation of incoherent pairs for B > $B_c$ and T > $T_c$[30]. Regardless of the origin, the correlations responsible for the pseudogap are clearly compatible with the onset of phase-coherent superconductivity.

**Quenching pairing and pseudogap with hBN**

Further insight regarding superconductivity and the pseudogap phase in MATBG is provided by studying MATBG aligned with hBN. Anecdotally, transport experiments do not report superconductivity in MATBG samples that are presumed to be well-aligned with hBN[31,32]. In examining the role of hBN alignment, STM studies are particularly advantageous, as they can directly visualize and distinguish the graphene-graphene (G-G) and graphene-hBN (G-hBN) moiré structures. Fig. 5b shows a set of representative topographic images of Device C, taken at different $V_s$ and $V_g$ in order to disentangle the different structural roles of the two moiré patterns (see SI). Surprisingly, these images show perfect alignment between the AA sites of the G-G moiré and the carbon-boron regions of the G-BN moiré. This suggests a propensity for MATBG aligned to hBN to undergo a moiré-scale incommensurate-commensurate transition when the two moiré length scales are similar. In the schematic in Fig. 5a, we label these substrate-modified AA sites as "AAb" sites to reflect this alignment configuration. Likewise, the AB/BA sites of MATBG are made inequivalent by the hBN, forming "ABa" ("BAa") regions where atoms in the top (bottom) graphene sheet are in register with atoms in the top hBN layer. This incommensurate-commensurate transition contrasts with the formation of a super-superlattice due to a long-wavelength interference between the two moiré patterns.

DT-STS and DT-PCS on Device C show that alignment with hBN dramatically alters the electronic properties of MATBG (Figs. 5c-e). In contrast to the semi-metallic behavior we observe in unaligned samples at the CNP, STS acquired at the center of an AAb site shows a gap (convolved with Coulomb charging effects) at the Dirac point due to sublattice symmetry breaking[33–35], and the resulting insulating behavior of MATBG at the CNP is directly probed using PCS (Figs. 5c and 5e). In agreement with transport studies[31,32], we also find correlated and Chern insulators at $\nu$ = +2 and +3, respectively (Fig. 5c; see also SI). In contrast with unaligned samples, aligned samples show neither a cascade of transitions nor evidence for superconductivity or the pseudogap, despite the twist angle of this device (1.08°) being near those with the maximal $T_c$ in transport measurements on unaligned devices[1,5]. Overall, DT-STS and DT-PCS show that hBN alignment is detrimental to the formation of both the pseudogap and the superconducting phases of MATBG, as evidenced by contrasting data in Figs. 1-2 with that of Fig. 5. Furthermore, the ability to identify superconductivity and a pseudogap phase in unaligned MATBG and their absence in MATBG aligned to hBN demonstrates the utility of our combined DT-STS and DT-PCS technique, since the existence of superconductivity in some flat-band materials[36] as well as the importance of $C_2T$ symmetry[37–39] is currently heavily contested.

**Discussion**

Cumulatively, our findings provide substantial evidence that pairing in MATBG is unconventional and distinct from that of a BCS mechanism. STS does not show an isotropic gap with a size consistent with that expected from a $T_c$ ~ 1.2 K s-wave BCS superconductor, but shows a V-shaped DOS consistent with that of a nodal superconductor, where the details of the spectra vary with twist angle and strain (see SI). The PCS measurements corroborate this picture and additionally show an unusual linear suppression of the Andreev excess current approaching $T_c$ (Fig. 3d). This behavior is similar to reports in other unconventional superconductors[40,41] and has been suggested to be related to pair-breaking effects due to inelastic scattering from bosonic modes. There are many candidates for bosonic modes in MATBG, ranging from phonons to more exotic collective isospin fluctuations[42]; however, a key ingredient for this scenario is the presence of a sign-changing order parameter, which makes scattering from such modes pair breaking[43]. As an aside, if pairing is spin-triplet in nature, the ratio of the enhanced conductance near zero bias to the background conductance in the Andreev spectra is incompatible with an equal-spin-pairing order parameter (see SI). Moreover, like the underdoped cuprate superconductors[27], MATBG shows contrasting behavior between the energy scales describing tunneling and Andreev reflection. Without further experiments, it is difficult to distinguish between different explanations for this dichotomy (i.e. a precursor broken-symmetry phase or preformed pairing without coherence[30]). Overall, the experiments presented here provide clear constraints for constructing a model of the pairing mechanism in this novel electronic material that lies beyond the BCS paradigm.

**Acknowledgements**

We thank P. Jarillo-Herrero, A. H. MacDonald, and S. A. Kivelson for helpful discussions. We thank C.-L. Chiu, G. Farahi, and H. Ding for helpful technical discussions. This work was primarily supported by the Gordon and Betty Moore Foundation's EPiQS initiative grants GBMF9469 and DOE-BES grant DE-FG02-07ER46419 to A.Y. Other support for the experimental work was provided by NSF-MRSEC through the Princeton Center for Complex Materials NSF-DMR- 2011750, NSF-DMR-1904442, ExxonMobil through the Andlinger Center for Energy and the Environment at Princeton, and the Princeton Catalysis Initiative. A.Y. acknowledge the hospitality of the Aspen Center for Physics, which is supported by National Science Foundation grant PHY-1607611, and Trinity College, Cambridge UK where part of this work was carried in with the support of in part by a QuantEmX grant from ICAM and the Gordon and Betty Moore Foundation through Grant GBMF9616. K.W. and T.T. acknowledge support from the Elemental Strategy Initiative conducted by the MEXT, Japan, grant JPMXP0112101001, JSPS KAKENHI grant 19H05790 and JP20H00354.

**Author Contributions**

M.O., K.P.N., D.W., and A.Y. designed the experiment. M.O., D.W., and K.P.N. fabricated the devices used for the study. M.O., K.P.N., D.W., and R.L.L carried out STM/STS and PCS measurements, with invaluable input from X.L. on the latter. M.O., D.W., and K.P.N. performed the data analysis. K.W. and T.T. synthesized the hBN crystals. All authors discussed the results and contributed to the writing of the manuscript.

**Figure Captions:**

**Figure 1 | Scanning tunneling spectroscopy of the tunneling gap of superconducting MATBG. a,** Schematic of the experimental set-up. MATBG, biased at $V_s$, sits atop hBN/SiO$_2$/Si, while $V_g$ is applied to Si to tune the carrier density. **b,** STM topographic image of MATBG. **c,** Tunneling d$I$/d$V(V_s, V_g)$ taken at the center of an AA site in Device A (1.13°, 0.4% strain) shows the conduction and valence flat bands pinned to E$_F$. The red dashed-line box highlights a set of gaps in the valence flat band. **d,** Higher-resolution d$I$/d$V(V_s, V_g)$ for Device A shows a gap at $\nu$ = -2 (CI; correlated insulator) and a gap between $\nu$ = -2 and $\nu$ = -3 (SC; superconductor). A line cut of d$I$/d$V(V_g)$ at $V_s$ = 0 V is shown on the right. **e,** d$I$/d$V(V_s)$ spectra for Device A at $V_g$ = -22.6 V (top) and $V_g$ = -25.8 V (bottom). **f,** Same as **d**, except for Device B (1.06°, 0.1% strain). **g,**

d$I$/d$V$($V_s$) spectra for Device B at $V_g$ = -19.8 V (top) and $V_g$ = -25.6 V (bottom). See SI for tunneling parameters.

**Figure 2 | Point-contact spectroscopy and Andreev reflection for MATBG. a,** Schematic of the Andreev reflection process measured using density-tuned point-contact spectroscopy (DT-PCS). The STM tip is brought into point contact with the surface of MATBG, and the two-terminal conductance G($V_s$, $V_g$) is measured. **b,** Line cut of point-contact G($V_g$) at $V_s$ = 0 V for Device A' (same as Device A, different region; 1.01° twist angle, 0.2% strain) at five magnetic field strengths between 0 T and 200 mT. Strong suppressions of G($V_g$) occur near $\nu$ = -2, +2, and +3 as a result of correlated insulating phases near these integer fillings (CI; red shaded bars). A dip in G($V_g$) occurs near charge neutrality (CNP; gray shaded bar). An enhancement of G($V_g$) occurs between $\nu$ = -2 and $\nu$ = -3 as a result of the excess current measured in the superconducting phase (SC; green shaded bar). Curves are vertically offset by the horizontal black lines for clarity. **c,** Line cut of point-contact G($V_s$) spectra at $V_g$ = -21 V, in the superconducting carrier-density range, at five magnetic-field strengths between 0 T and 200 mT. Curves are offset for clarity. **d,** Line cut of point-contact G($V_s$) spectra at $V_g$ = -21.8 V, in the superconducting carrier-density range, at sixteen temperatures between 300 mK and 1.3 K. **e,** Point-contact G($V_s$, $V_g$) and dG/d$V_s$($V_s$, $V_g$) for different values of the out-of-plane magnetic field showing the disappearance of Andreev reflection at around 50 mT. **f,** Point-contact G($V_s$, $V_g$) and dG/d$V_s$($V_s$, $V_g$) for different values of the temperature showing the disappearance of Andreev reflection at around 1.3 K. **g,** Side-by-side tunneling d$I$/d$V$($V_s$, $V_g$) into an AA site and point-contact G($V_s$, $V_g$) in the same location in Device A'. Gaps observed in tunneling marked as CI coincide with highly resistive states in G($V_s$, $V_g$), while the tunneling gap marked as SC coincides with Andreev reflection. See SI for tunneling and PCS parameters.

**Figure 3 | Tunneling and Andreev reflection spectra curve fits. a,** Dynes-function fits to the experimental tunneling spectrum (blue curve) measured at $V_g$ = -25.8 V for Device A at 200 mK and B = 0 T using the model quasiparticle DOS for a nodeless s-wave superconductor with all free parameters (red curve) and with fixed lifetime broadening parameter $\Gamma$ = 0.07 meV (gray curve) (see SI for details). **b,** Same as **a,** except using the model quasiparticle DOS for a nodal superconductor (e.g. p-, d-, f-wave). **c,** Andreev reflection spectra (solid curves) obtained in Device A' at $V_g$ = -21.8 V at fifteen temperatures between 300 mK and 1 K, fit with the BTK model (dashed curves) with fixed barrier transparency parameter Z = 0.1. **d,** Excess current $I_{exc}$ and the superconducting energy gap $\Delta_{AR}$ extracted from the BTK fits in **c** (see SI for details).

The excess current shows an anomalous linear dependence on the temperature, indicative of unconventional superconductivity. **e,** Tunneling d$I$/d$V$($V_s$) spectra acquired at the center of an AA site between $V_g$ = -20 V and $V_g$ = -26 V in Device A'' (same as Device A, different region; 0.99° twist angle), measured at T = 4.1 K. Curves are offset by 0.6 nS for clarity. A suppression of the DOS near $E_F$ is observed at temperatures above $T_c$ for superconductivity observed at -3 < $\nu$ < -2 (See SI for full d$I$/d$V$($V_s$, $V_g$) for all observations of the high-temperature pseudogap phase). See SI for tunneling parameters.

**Figure 4 | Pseudogap regime and phase diagram of hole-doped MATBG. a,** Tunneling d$I$/d$V$($V_s$) spectra at $V_g$ = -25 V taken at the center of an AA site in Device A at $B_\perp$ = 0 T (red curve), 0.5 T (purple curve), and 1 T (blue curve), which show the persistence of a prominent gap at $E_F$ well above $B_c$ for MATBG. **b,** Tunneling d$I$/d$V$($V_s$, $V_g$) and d$I$/d$V$($V_s$) spectra on an AA site in Device B for $V_g$ = -19 V to $V_g$ = -34 V and for $B_\perp$ = 0 T, 3 T, and 6 T. Curves are offset by 7.5 nS for clarity. At $B_\perp$ = 0 T, a gap opens and closes near $\nu$ = -2 due to the correlated insulating (CI) phase, followed by a gap for the superconducting (SC) phase at -3 < $\nu$ < -2. At $B_\perp$ = 3 T, the gap at -3 < $\nu$ < -2 is a result of the pseudogap (PG) regime. At $B_\perp$ = 6 T, a series of large gaps appear that correspond to correlated Chern insulating (ChI) phases with Chern numbers C = -3, -2, -1. See SI for tunneling parameters. **c,** A proposed schematic phase diagram for MATBG as a function of flat-band filling factor $\nu$ and magnetic field $B_\perp$ in the hole-doped regime. ($\nu_{LL}$ is Landau-level filling factor.) Near –3 < $\nu$ < -2, we observe an unconventional superconducting phase at low magnetic fields, which transitions into a pervasive pseudogap regime at high magnetic fields.

**Figure 5 | DT-STS and DT-PCS on non-superconducting MATBG aligned to hBN. a,** STM topographic image of MATBG that is perfectly commensurate with the underlying hBN substrate. Atomistic schematics show the stacking configurations of carbon, boron, and nitrogen for different regions of the moiré pattern. **b,** STM topographic images of MATBG aligned to hBN for different values of $V_s$ and $V_g$, highlighting the graphene (G-G) moiré pattern and the graphene-hBN (G-hBN) moiré pattern. **c,** Side-by-side comparison of tunneling d$I$/d$V$($V_s$, $V_g$) into an AAb site and point-contact G($V_s$, $V_g$) for Device C (1.08° G-G twist angle, 0.1% G-G interlayer strain, 0.5 ± 0.1° G-hBN twist angle), which uses a graphite gate instead of a silicon

gate. No signatures of a superconducting gap, pseudogap, or of Andreev reflection can be seen in either measurement. **d,** d$I$/d$V$($V_s$) spectra from **c**, offset by 15 nS (left) and 20 nS (right) for clarity. **e,** Tunneling d$I$/d$V$($V_g$) and PCS G($V_g$) line cuts from **c** for $V_s$ = 0 V. See SI for tunneling and spectroscopy parameters.

## Methods

### STM measurements

STM/STS measurements were performed on a homebuilt dilution-refrigerator STM[15] with tungsten tips prepared on a Cu(111) surface. The carrier density in MATBG was tuned by a gate voltage $V_g$ applied to Si (or a graphite gate for Device C), while $V_s$ is applied to the sample. d$I$/d$V$ is measured through lockin detection of the AC tunnel current in response to an AC modulation $V_{rms}$ added to $V_s$. Initial tunneling parameters for STS are chosen to avoid phonon-induced inelastic tunneling[44].

We used two experimental protocols for avoiding unwanted local gating from the tip[20]. First, we used an STM tip that has been freshly prepared (field emission, pulsing, poking) and calibrated on a cleaned single-crystal metal, paying particular attention to protecting the tip from polymer residue contamination that often lies on the surface of two-dimensional (2D) material devices. Second, we use an STM tip and metal crystal that are made of materials (e.g. tungsten and copper) that are work-function-matched with graphene. Careful preparation of the tip and sample are essential because when polymer residue on the device's surface attaches to the tip, spectroscopic features of the tunnel junction are compromised, and topographic images often show "drag patterns" caused by the motion of a particle in the tunnel junction or by flexing of the tip apex[45–47]. Since these drag patterns may be misinterpreted as tip-induced strain effects, we provide evidence of our clean and stable tip-sample junctions in Fig. S16 in the Supplementary Information. Fig. S16 shows two topographic images without a drag pattern that are essentially identical despite a three-orders-of-magnitude change in the junction resistance.

### PCS measurements

PCS measurements were performed by moving the STM tip a few nanometers (relative to the tip height during tunneling) into the MATBG surface. This does not damage the graphene. Differential conductance G($V_s$, $V_g$) is then measured through lock-in detection of the AC current in response to an AC modulation $V_{rms}$ added to $V_s$, while dG/d$V_s$($V_s$, $V_g$) is simply the numerical derivative of the measured G($V_s$, $V_g$). We note that the conductance G($V_s$), appears to be slightly suppressed around zero bias in the metallic state of MATBG at millikelvin temperatures, but this suppression vanishes at T = 1.3 K. Since this suppression is present at all $V_g$ and at magnetic fields above $B_c$, we conjecture that this is due to non-Ohmic contact, possibly between the graphene and the Ti/Au electrodes. When MATBG is superconducting, the finiteness of the critical current and the proximity effect may also contribute to the suppression of the conductance around the Andreev peaks[48]. See the Supplementary Information for more details on the PCS measurements.

The data in Figs. 2b, c, e, and f were acquired together, and the data in Figs. 2d and g were acquired together. Between these two sets of data, the tip was withdrawn from the surface, and then point contact was re-established in the same location. The temperature-dependent data in Fig. 2d was acquired by heating the $^3$He-$^4$He mixture to T = 1.3 K and then measuring PCS as the dilution refrigerator is cooled. The temperatures in Fig. 2d are measured via a $RuO_2$ thermometer in the STM head. The tip likely drifts relative to the sample during this measurement.

Since Yankowitz *et al.*[2] has shown that superconductivity in twisted bilayer graphene can be tuned with pressure, we examined the role of tip-induced pressure/strain during a PCS measurement. Fig. S6 in the Supplementary Information shows tip-height-dependent PCS, showing that the energy scale for Andreev reflection $\Delta_{AR}$ is unchanged as the tip is pressed further into MATBG. This, along with the fact that the density range, $T_c$, and $B_c$ of superconductivity in PCS match those of transport experiments, verifies the one-to-one correspondence of STS and PCS at the same location. See Section D of the Supplementary Information for a further discussion.

**Sample preparation**

Devices were fabricated using a "tear-and-stack" method[49] in which a single graphene sheet is torn in half by van der Waals interaction with hBN. The two halves are rotated relative to each other and stacked to form MATBG. As Device B is Device A from Ref. [20], a full description of the fabrication procedure can be found therein. To summarize, graphene and hBN are picked up with polyvinyl alcohol (PVA). Then, to flip the heterostructure upside down, the heterostructure is pressed against an intermediate structure consisting of polymethyl methacrylate (PMMA)/transparent tape/Sylgard 184, and the PVA is dissolved via water injection. The heterostructure is then transferred to an $SiO_2$/Si chip with pre-patterned Ti/Au electrodes. Residual polymer is dissolved in dichloromethane (DCM), water, acetone, and isopropyl alcohol (IPA). This chip is annealed in ultra-high vacuum (UHV) at 170 °C overnight and 400 °C for 2 hours. Device A is prepared in a similar manner, except the PMMA is replaced with Elvacite 2550, and N-Methyl-2-pyrrolidone (NMP) is added as a solvent. For Device C, the intermediate structure consists only of Sylgard 184 on a glass slide, and a graphite gate is added to the heterostructure.

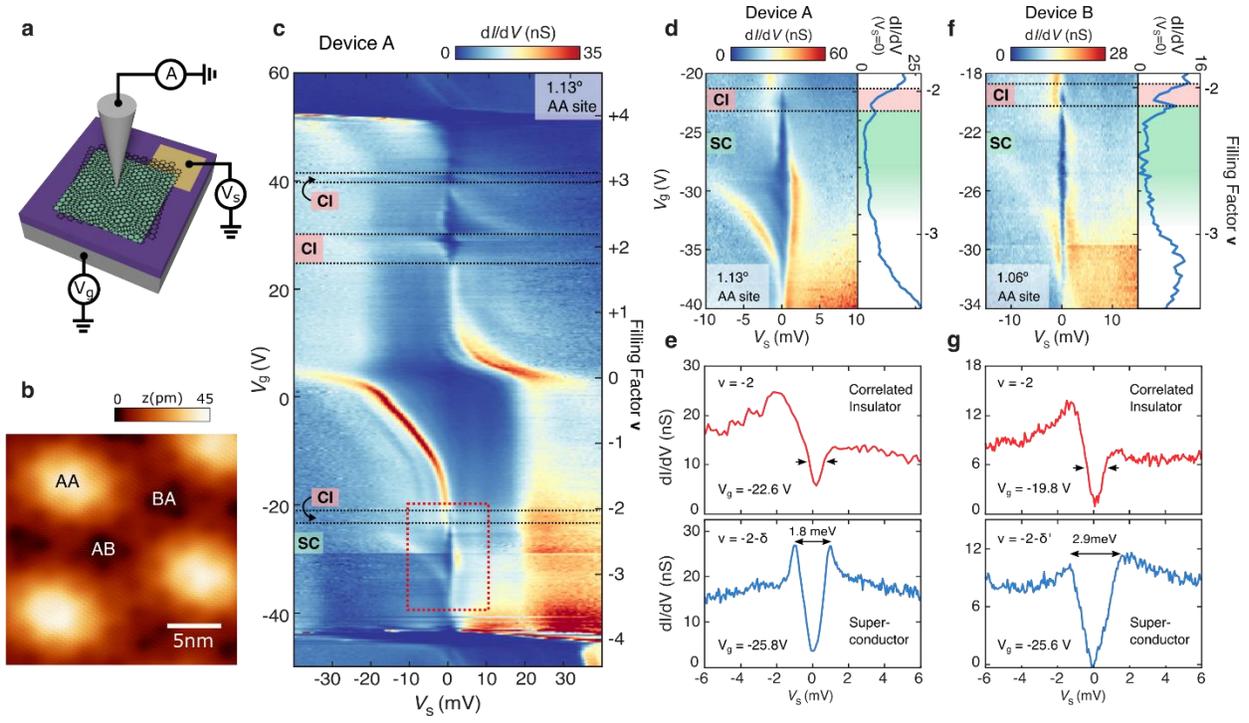

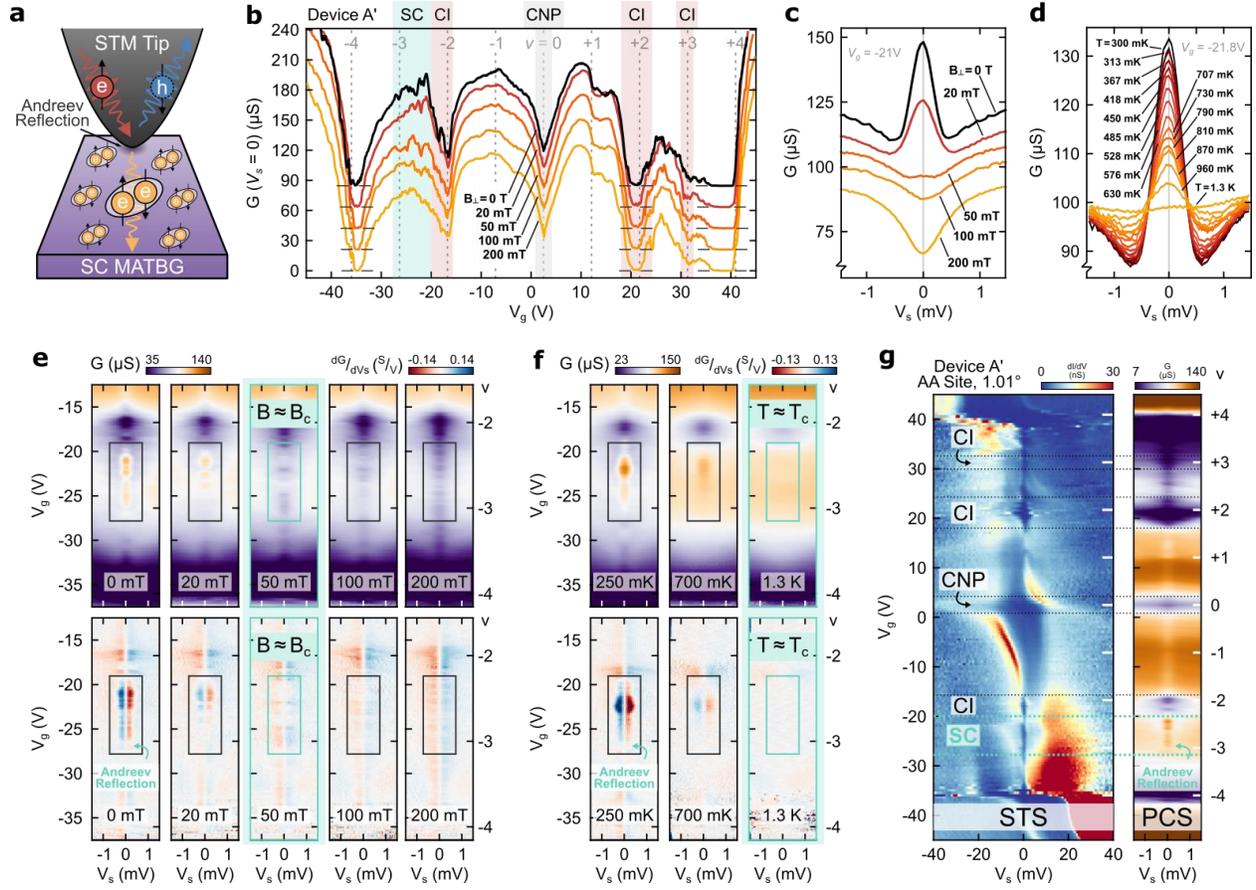

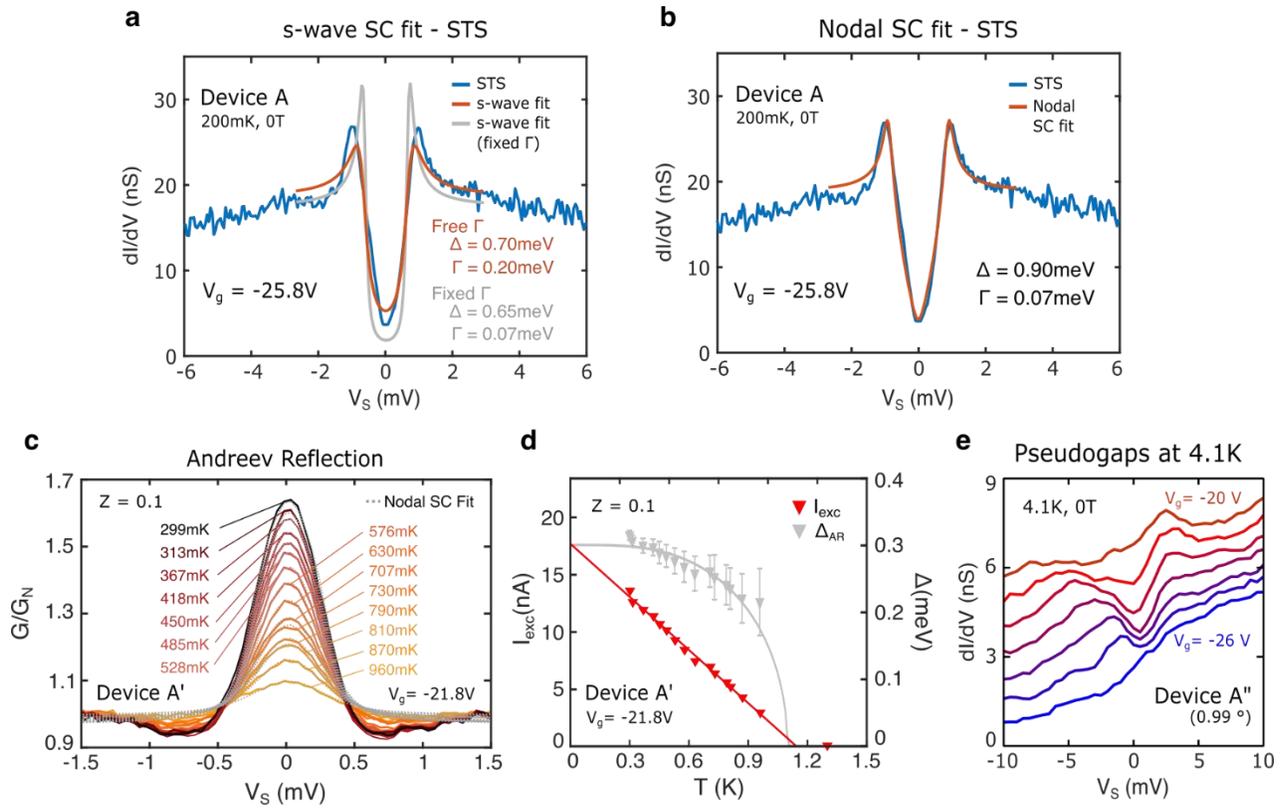

Figure 4

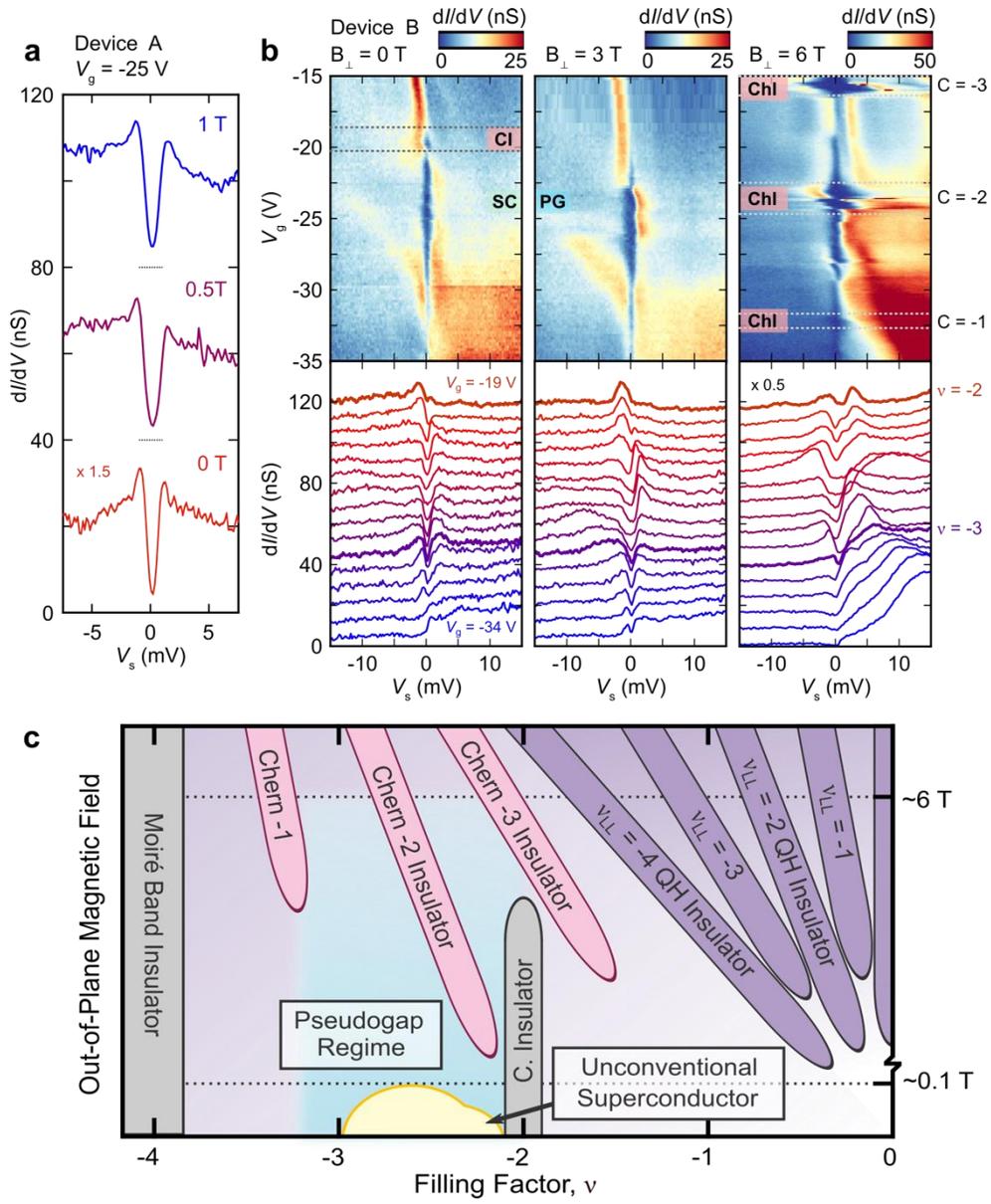

Figure 5

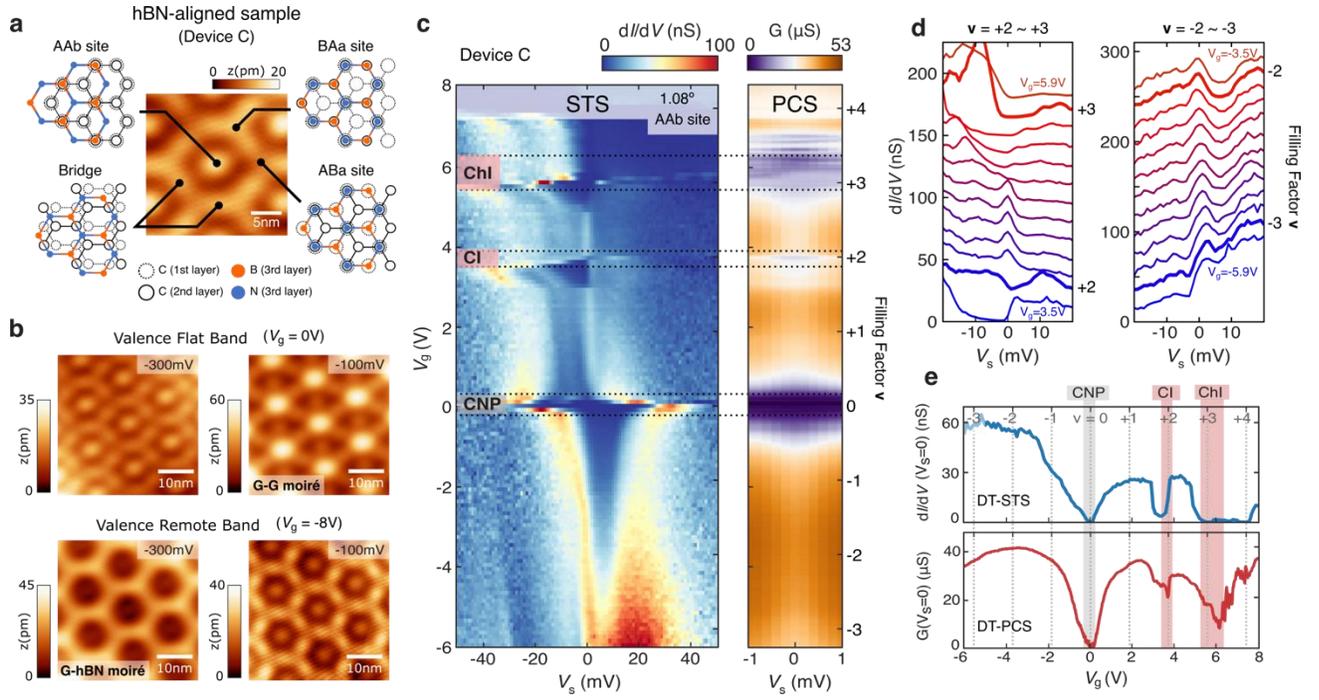

# Supplementary Information: Evidence for an Unconventional Mechanism of Superconductivity in MATBG

Table of contents:

## A. Additional tunneling and Andreev spectroscopy measurements between $\nu$ = -2 and $\nu$ = -3

In the main text, we discussed the presence of a V-shaped tunneling gap that appears at the same density as an Andreev reflection spectrum acquired at the same location in an unaligned MATBG sample. We observed this gap at millikelvin temperatures five times in four devices, showing slight variations in shape and size. However, none of these observed gaps take the shape of an isotropic s-wave superconducting gap. Fig. S1 shows d$I$/d$V$($V_s$, $V_g$) displaying the $\nu$ = -2 gap and the -3 < $\nu$ < -2 gap for Device E (angle : 1.10°) and Device F (angle : 1.10°, strain : 0.1%).

The set of 5 total observations of this STS phenomenology in the valence flat band (two depicted in main text Fig. 1, one depicted in main text Fig. 2, and two depicted in Fig. S1) shows similar realizations of the same basic phase diagram of MATBG (i.e. a correlated insulator near

$\nu$ = -2 and a superconductor between $\nu$ = -2 and $\nu$ = -3). Theoretically, interlayer strain, which is practically unavoidable among devices fabricated with any variation of the classic "tear-and-stack" method, is thought to strongly influence coupling in various pairing channels of different symmetries[1], which may explain some of the more detailed difference in spectroscopy in each sample. However, our ability to topographically disentangle the roles of the twist angle, the interlayer relative strain, and the hBN alignment allows for us to identify the most robust spectroscopic signatures of unconventional superconductivity in MATBG.

In Fig. S2, we show DT-PCS measurements in Device D that show Andreev reflection. Using BTK-model fits (Fig. S2c; see also Section E) for a nodal superconducting order parameter, the Andreev gap in this device reaches an optimal value of $\Delta_{AR}$ = 0.71 meV at filling $\nu$ = -2.44. In contrast to the data from Device A' in main text Fig. 2, this data shows a suppression of the zero-bias conductance, which is common in PCS junctions with larger barrier strength $Z$.[2]

## B. Analysis of the tunneling gaps & possible pairing symmetries for -3 < $\nu$ < -2

To capture the tunneling spectra, we compare our data to that of the density of states (DOS) calculated using a modified BCS model, which includes the effect of quasiparticle lifetime, as introduced by Dynes[3], as well as the possibility of anisotropic and nodal gaps. The Dynes model for the DOS for an isotropic s-wave superconductor is given by:

$$\rho(E) = \rho_N Re\left[\frac{|E - i\Gamma|}{\sqrt{(E - i\Gamma)^2 - \Delta^2}}\right]$$

where the quasiparticle broadening $\Gamma$ and the superconducting gap size $\Delta$ can be free fitting parameters. We model the background normal-state DOS $\rho_N$ as either a constant or a linear function of energy. To include the possibility of a nodal gap in this model, we introduce the angular-dependent factor

$$\Delta(\theta) = \Delta_0 \cos(l\theta)$$

where $l$ is the orbital angular momentum quantum number for the pair wavefunction (i.e. $l = 1, 2, 3$ for p-, d-, and f-wave superconducting order parameters, respectively)[4]. The energy dependence of the DOS for a nodal superconductor is obtained by integrating the DOS at a given energy over all angles[4]:

$$\rho(E) = \frac{\rho_N}{2\pi}\int_0^{2\pi} d\theta \, Re\left[\frac{E - i\Gamma}{\sqrt{(E - i\Gamma)^2 - \Delta(\theta)^2}}\right]$$

For nonzero angular momentum scenarios ($l > 0$), we can simplify the equation to the following form:

$$\rho(E) = \frac{\rho_N}{2\pi} \int_0^{2\pi} d\theta \, Re\left[\frac{E - i\Gamma}{\sqrt{(E - i\Gamma)^2 - \Delta_0^2 \cos^2(l\theta)}}\right] = \frac{\rho_N}{2\pi} \int_0^{2\pi} d\theta \, Re\left[\frac{E - i\Gamma}{\sqrt{(E - i\Gamma)^2 - \Delta_0^2 \cos^2\theta}}\right]$$

Thus, the tunneling conductance measured between the STM tip and a superconducting sample is

$$\frac{dI}{dV}(V) \propto \int_{-\infty}^{\infty} dE \left.\frac{df(\varepsilon)}{d\varepsilon}\right|_{\varepsilon = E - eV} \rho(E) \rho_{tip}$$

where $\rho_{tip}$ is the tip DOS, which we assume to be an energy-independent constant. In Supplementary Section E below, where we discuss Andreev reflection, we will extend this discussion to include the spin orientation of Cooper pairs and its impact on tunneling and Andreev spectroscopy.

The main text Figs. 3a,b show the nodeless s-wave and nodal Dynes-formula fitting analyses for the gap observed at $V_g$ = -25.8 V in Device A (this gate voltage was chosen because it has the least asymmetric background as the van Hove singularity shifts through $E_F$). Using a least-squares regression procedure, we obtain the best fit using $\Delta = 0.90$ meV for the nodal model (red curve).

The nodal-superconductor fit closely matches our experimental results, aptly capturing the sharp V-shaped dip in the spectrum near zero bias and the sloped coherence-peak structures observed at the edges of the tunneling gap. In contrast, the nodeless s-wave fits poorly capture these two aspects of the observed tunneling gap. The quasiparticle lifetime broadening required for the best fit to the nodeless s-wave model (red curve in main text Fig. 3a; $\Gamma$ = 0.20 meV) is much larger than that required for the nodal case ($\Gamma$ = 0.07 meV). Generally, one expects s-wave superconductors to be more robust against impurity scattering effects than superconductors with sign-changing order parameters, making the larger $\Gamma$ for the s-wave scenario anomalous (see for comparison, the superconducting gap of Al(100) in Fig. S3c). We have found no evidence for impurity scattering in our measurements to justify the large broadening needed for the s-wave model. In main text Fig. 3a (gray curve), we show the fit to our experimental data using the isotropic Dynes formula for $\Gamma$ fixed to the same lifetime broadening as used for the fit to the nodal model ($\Gamma$ = 0.07 meV). The comparison between main text Fig. 3a (both nodeless fits) and main text Fig. 3b (the nodal fit) clearly shows the superiority of the nodal model in capturing the features of our data.

We also performed the same analysis on the tunneling spectrum in Device B and Device F (Figs. S3a and S3b). In both cases, the nodal-superconductor fits capture the features in the experimental curves better than for the s-wave fits. As with Device A, fits to the s-wave model for the data from these devices require large lifetime broadening parameters ($\Gamma \sim 0.3 \Delta$).

We note that the electron-hole symmetry present in the lower panels of main text Figs. 1e,g is not present throughout the gate-voltage range of the gap. When analyzing spectroscopy as a function of gate voltage in both main text Fig. 1d from Device A and main text Fig. 1f from Device B, we observe a van Hove singularity (vHs) below the Fermi level at fillings $\nu$ > -2, appearing at negative $V_s$. At fillings near $\nu$ = -3, this vHs appears at positive $V_s$, having passed through the Fermi level between $\nu$ = -2 and $\nu$ = -3. This high DOS feature is responsible for causing electron-hole asymmetry in the normal-state DOS, upon which the tunneling gap is superimposed at these fillings.

In main text Fig. 3a, we chose to analyze a spectrum that is electron-hole symmetric out of convenience: we can fit this spectrum to a modified BCS DOS spectrum without needing to make assumptions about the normal-state DOS. At other gate voltages, the spectrum still appears V-shaped, but has significant electron-hole asymmetry that makes the fitting analyses less straightforward. Nevertheless, we can still do fitting analyses despite this electron-hole asymmetry by fitting only the low-energy portion of the spectroscopic gap (Fig. S4). A similar analysis to ours was used in STS studies of the high-temperature cuprate superconductors, where electron-hole asymmetry is consistently seen[5]. In Fig. S4, the nodal fits are consistently superior to the s-wave fits, with each nodal fit having a lower $\chi^2$ value than the s-wave fit to the same spectrum.

## C. Locality of the DT-PCS method

DT-PCS is a highly spatially local method that is most sensitive to the region directly underneath the tip. Here we present experimental observations and an idealized model that support the idea that regions far away from the tip do not contribute significantly to the DT-PCS signal.

First, the side-by-side STS-PCS of Device A' in main text Fig. 2g shows an unambiguous correspondence between spectroscopic gaps in tunneling d$I$/d$V$($V_g$) and resistance features in point-contact G($V_g$). This remarkable correspondence occurs despite significant spatial inhomogeneity in Device A', from which measurements in main text Fig. 1c (1.13°), main text Fig. 2 (1.01°), and main text Fig. 3e (0.99°) are derived. A Nomarski differential interference contrast (DIC) image of Device A' in Fig. S5b shows a sample with an area that is roughly 4400 µm² (the twisted graphene covers the entire hBN substrate). If PCS was a measurement that averaged over the entire sample, the correspondence between STS and PCS seen in Fig. 2g of the main text would be absent.

Second, Figs. S5c,d show two PCS G($V_g$) data sets at $V_s$ = 0 V, obtained roughly 200 nm away from each other in Device C. These two data sets show different transport phenomenology near $\nu$ = -1. The electronic structure of moiré heterostructures is highly sensitive to properties such as twist angle, strain, and moiré-commensurability that can spatially vary across a sample. Here we have demonstrated that PCS can detect an insulating state at $\nu$ = -1 in the region probed in Fig. S5d while not detecting an insulating state only 200 nm away in Fig. S5c.

Third, we can understand the locality of DT-PCS by considering an idealized model on an azimuthally symmetric sample (schematic diagram in Fig. S5a). While this model is idealized, it nevertheless provides a reasonable explanation for why regions far from the tip have little influence on the PCS signal. Consider a two-dimensional, circular sample of radius $R$ centered at the origin (0,0) and an STM tip touching the origin with point-contact radius $R_0 \ll R$. Consider a voltage $V$ applied to the metallic STM tip, while the outer edge of the sample is held at ground $V = 0$. What is the voltage drop across a sample of resistivity $\rho(r, \theta) \equiv \rho(r)$ as a function of radius?

Kirchoff's current law requires that the total current passing through every radial ring around the origin must be the same. Thus, for a radially symmetric current density distribution $J(r)$ at two radii $r_1$, $r_2$, we know that

$$I_1 = r_1 \int_0^{2\pi} J(r_1) d\theta = r_2 \int_0^{2\pi} J(r_2) d\theta = I_2 \, .$$

Using Ohm's law ($E(r) = \rho(r) J(r)$), we find a relationship between the electric field at these two radii:

$$\frac{r_1 E(r_1)}{\rho(r_1)} = \frac{r_2 E(r_2)}{\rho(r_2)} \, .$$

Thus, the voltage at a point in the sample at radius $r$ is as follows:

$$\Delta V \equiv V(R_0) - V(r) = -\int_{R_0}^{r} E(r') dr' = -\frac{R_0 \, E(R_0)}{\rho(R_0)} \int_{R_0}^{r} \frac{\rho(r')}{r'} dr' \, .$$

The radial factor in the denominator of the integral is key to the locality of this technique. Regions near the tip (i.e. small $r'$) contribute the most to this integral.

## D. Method and accuracy of DT-PCS measurements
### i. Additional details for DT-PCS method

To perform PCS measurements using the STM, the STM tip is first placed within tunneling range of the sample using a closed current feedback loop configuration (setpoint parameters: $V_s$ = -98 mV, $I$ = 10 pA). The back gate of the sample is set to a voltage for which the MATBG sample is metallic. We often use a very large negative gate voltage (e.g. for a Si-gated device, $V_g$ ~ -60 V), which places the chemical potential of MATBG deep within the metallic remote band below the flat bands. This will be used later to determine whether or not tip-sample contact has been made. After some time (~10 minutes), once the piezoelectric scanner drift has reduced significantly, the current feedback loop is opened and the sample voltage is reduced to 1 - 2 mV. Then, the tip is manually displaced in the z-direction by a few nanometers, in increments of one angstrom at a time, until the current $I$ increases to a value of tens of nanoamps and is stable over several minutes. The exact value of $I$ is logged and used as a benchmark for achieving the same transparent barrier conditions for PCS measurements in different locations with the same tip.

### ii. Accuracy of DT-PCS

For tunneling and PCS measurements, we use a passive voltage divider (1:100 for the DC component of $V_s$ and 1:1000 for the $V_{rms}$ lock-in modulation) for increasing the energy resolution by reducing the voltage noise from voltage sources. In the circuit diagram depicted in Fig. S7a, we use resistors for the voltage divider. For tunneling measurements, since the tip-sample junction resistance $R_J$ is on the order of gigaohms, the voltage-divider output is well-approximated by

$$V_{out} \approx \frac{R_2}{R_1 + R_2} V_{source},$$

for $R_2$ = 1 kΩ. For the PCS measurement, however, the junction resistance $R_J$ becomes comparable to $R_2$, so the voltage across the total device resistance $R_T$ (contact resistance $R_C$ + junction resistance $R_J$) is

$$V_T = \frac{R_2 R_T}{(R_1 + R_2)(R_T + R_{cable}) + R_1 R_2} V_{source},$$

where $R_{cable}$ is the cryostat cable resistance (there are two 200 Ω stainless steel cables to and from the device in the dilution refrigerator).

Comparing $V_T$ to the voltage expected from a simple voltage divider

$$V_s = \frac{R_2}{R_1 + R_2} V_{source},$$

the voltage error in the limit $R_1/R_2 \gg 1$ is as follows:

$$Error = \frac{V_S - V_T}{V_S} = \frac{(R_1 + R_2)R_{cable} + R_1 R_2}{(R_1 + R_2)(R_T + R_{cable}) + R_1 R_2} \approx \frac{R_{cable} + R_2}{R_T + R_{cable} + R_2}.$$

Fig. S7b analyzes the $R_T$-dependence of this voltage error. In our measurements, $R_T$ is between 10 kΩ ~ 25 kΩ, indicating that the Andreev gaps $\Delta_{AR}$ may be overestimated by as much as ~10%.

### iii. Tip-induced pressure and strain

Pressure[6] and strain[7] influence the electronic properties of twisted bilayer graphene. For example, Yankowitz *et al.*[6] showed that GPa pressures are able to induce superconductivity in a non-magic-angle sample (1.27°) that did not superconduct at zero pressure. In principle, because the narrow magic-angle condition arises from fine-tuning of the energies of the interlayer and intralayer hopping processes[8], it is possible that applied pressure converts a superconducting magic-angle sample into a sample without superconductivity. Hence, it is important to consider whether pressure and strain induced by the tip during a PCS measurement affects the electronic properties of MATBG. In our study, we can identify that our sample is magic-angle through STM topography and spectroscopy, and yet the point-contact measurement still shows Andreev reflection, indicating that pressure from the tip did not extinguish superconductivity in our devices. Moreover, Fig. S6a shows that the Andreev energy scale does not change even when we push the STM tip 5 nm deeper into the sample.

We note that the remarkable correspondence between DT-STS and DT-PCS in main text Fig. 2g seems to indicate that pressure and strain from the tip does not have a significant effect on the overall phase diagram of our device. This can be explained by the fact that the easily deformable STM tip (the apex of which is mostly copper) bends into the surrounding empty space as the tip is pushed into the sample.

Mesple *et al.*[9] documents the influence of tip-induced strain in twisted bilayer graphene in the tunneling regime. Figs. 3a, b, and c in Mesple *et al.* show a series of topographic images at different tunnel resistances (1.75 GΩ, 0.55 GΩ, and 0.15 GΩ) to demonstrate the impact of bringing the tip closer to the surface. Under our tip conditions (see Methods for information on how we prepare our STM tip), we do not observe the tip-induced strain reported by Mesple *et al.* Fig. S16 shows topographic images at tunneling resistances 20 GΩ and 0.02 GΩ that are essentially identical-looking.

## E. Blonder-Tinkham-Klapwijk (BTK) fitting of the Andreev spectrum between $\nu$ = -2 and $\nu$ = -3

To fit PCS spectra, we employ a modified version of the Blonder-Tinkham-Klapwijk (BTK) fitting function, which can also include the possibility of an anisotropic superconducting order parameter. Because the Andreev reflection process is sensitive to the phase of the superconducting order parameter, the propagation direction of impinging electrons upon a normal-superconductor junction, with respect to the crystallographic axes of the superconductor, greatly affects the theoretically predicted PCS spectrum. MATBG is a 2D system, and the tip-sample junctions in PCS are only made along the axis that is normal to the plane (the so-called "c-axis"). This forbids contributions of any sign-changing Andreev reflection processes towards the measured Andreev spectrum[10].

Andreev reflection spectra of an anisotropic superconductor are characterized by the BTK formula[10,11], which gives the normalized, dimensionless conductance $\sigma_R(E)$ at the quasiparticle energy $E$ as a function of the quasiparticle lifetime broadening $\Gamma$ at a given incident angle $\theta$ of an electron:

$$\sigma_R(E) = \frac{1 + \tau_N|\Gamma_+|^2 + (\tau_N - 1)|\Gamma_+\Gamma_-|^2}{|1 + (\tau_N - 1)\Gamma_+\Gamma_- \exp(i\varphi_- - i\varphi_+)|^2} \ .$$

Here $\Gamma_\pm = \frac{(E - i\Gamma) - \Omega_\pm}{|\Delta_\pm|}$, and $\Omega_\pm = \sqrt{(E - i\Gamma)^2 - |\Delta_\pm|^2}$. $\Delta_+$ and $\Delta_-$ are effective pair potentials, which are felt by hole-like quasiparticles and electron-like quasiparticles, respectively, and $\varphi_+$ and $\varphi_-$ are their corresponding phases. We use the 1D-BTK limit, where the barrier transparency parameter contains no geometrical factors:

$$\tau_N = \frac{1}{1 + Z^2} \ .$$

In the c-axis PCS analysis, $\Delta_+ = \Delta_-$, $\varphi_+ = \varphi_-$ and $\Gamma_+ = \Gamma_-$. In the nodeless s-wave case (Fig. S9b, left panel), the order parameter is isotropic, so pair potentials become $\Delta_+ = \Delta_- = \Delta_0$. In the nodal d-wave case (Fig. S9b, right panel), the order parameter is anisotropic, so the effective pair potential is $\pi$-periodic in the azimuthal plane:

$$\Delta_+ = \Delta_- = \Delta_0 \cos(2\phi).$$

The total conductance is obtained via integration over azimuthal angles:

$$\sigma(E) = \frac{\int_0^{2\pi} \sigma_R \tau_N d\phi}{\int_0^{2\pi} \tau_N d\phi} \ .$$

For the spin-triplet case, we consider a 2 x 2 matrix form for the gap function[12,13]:

$$\hat{\Delta}(k) = \begin{pmatrix} \Delta_{\uparrow\uparrow}(k) & \Delta_{\uparrow\downarrow}(k) \\ \Delta_{\downarrow\uparrow}(k) & \Delta_{\downarrow\downarrow}(k) \end{pmatrix}.$$

In an equal-spin-pairing (ESP) spin-triplet (or ferromagnetic) superconductor, $\Delta_{\uparrow\uparrow}(k)$ or $\Delta_{\downarrow\downarrow}(k) = \Delta_0(k)$ and $\Delta_{\uparrow\downarrow}(k) = \Delta_{\downarrow\uparrow}(k) = 0$. In this situation, the DOS of one spin population is given by the Bogoliubov-quasiparticle DOS, while the DOS of the other spin population is given by the normal-state DOS.

In an ESP superconductor, the tunneling conductance in the gap is due to the existence of the normal-state tunneling channel. This makes the conductance minimum no smaller than half of the normal-state DOS. By the same analogy, Andreev reflection occurs only for electrons of one spin-type, so the maximum conductance of the Andreev reflection spectrum in the gap cannot exceed 1.5 times the normal-state conductance. This is in contrast to the zero-bias conductance doubling expected for spin-singlet superconductors.

In main text Figs. 1e and 1g, the zero-bias conductance of the tunneling gap is much lower than half of the background conductance, while in the PCS spectroscopy in Fig. S9c, the maximum conductance of the normalized conductance is more than 50% larger than the background conductance. These observations rule out an ESP triplet superconductor scenario in MATBG. Thus, in the spin-triplet case, we use the opposite spin pairing (OSP) gap function,

$$\hat{\Delta}(k) = \begin{pmatrix} 0 & \Delta(\theta) \\ \Delta(\theta) & 0 \end{pmatrix}.$$

The dimensionless conductance $\sigma_{R\uparrow,\downarrow}(E)$ for each spin and the total conductance $\sigma(E)$ are[12]

$$\sigma_{R\uparrow}(E) = \sigma_{R\downarrow}(E) = \frac{1 + \tau_N |\Gamma_+|^2 + (\tau_N - 1)|\Gamma_+\Gamma_-|^2}{|1 + (\tau_N - 1)\Gamma_+\Gamma_- \exp(i\varphi_- - i\varphi_+)|^2},$$

$$\sigma(E) = \frac{\int_0^{2\pi}(\sigma_{R\uparrow} + \sigma_{R\downarrow})\tau_N d\phi}{\int_0^{2\pi} 2\tau_N d\phi}.$$

We use the gap function $\Delta_+ = \Delta_- = \Delta_0 \cos\phi$ for the p-wave case and $\Delta_+ = \Delta_- = \Delta_0 \cos(3\phi)$ for the f-wave case. The functional form becomes analytically equal to that of the d-wave case for a c-axis PCS measurement. Since the d-wave Dynes function fully captures the tunneling spectrum calculated from a d-wave BTK theory, the p- or f-wave Dynes function used in Supplementary Section B also provide a reasonable model to analyze the tunneling data.

At finite temperature, the point-contact conductance of the superconducting state $\sigma_S(V)$ can be represented by:

$$\sigma_S(V) = \sigma_N \int_{-\infty}^{\infty} dE \left.\frac{df(\varepsilon)}{d\varepsilon}\right|_{\varepsilon=E-eV} \sigma(E).$$

We compensate for the normal-state tip-sample resistance and non-ohmic electrode-sample contact resistances (or 'spreading resistance') by normalizing each spectrum using data obtained when the sample is in the normal state at B = 0.5 T > $B_c$. The normalized conductance can be represented as:

$$\frac{G(V)}{G_N(V)} = \frac{\sigma_S(V)}{\sigma_N(V)} \cdot \frac{\sigma_N + \sigma_C}{\sigma_S + \sigma_C},$$

where $\sigma_S$, $\sigma_N$, $\sigma_C$ and $G$ are the tip-sample junction conductance in the superconducting state, the tip-sample junction conductance in the normal state, sample-electrode contact conductance, and the total (contact + junction) conductance, respectively.

We can acquire the normalized point-contact conductance $\sigma_S/\sigma_N$ and rule out contact conductance effects (that is, omitting the second multiplicative factor above) when the contribution of the contact conductance is dominant in the second multiplicative factor above.

We show the tip-sample junction conductance to be much smaller than the sample contact conductance by using the normalized zero-bias conductance as a direct indicator of the quality of our sample contact. In the limit of a perfectly transparent tip-sample junction ($Z = 0$) at T = 0 and $\Gamma = 0$,

$$\sigma_S(V = 0)/\sigma_N(V = 0) = 2.$$

Experimentally, we find a smaller value for this ratio:

$$\sigma_S(V = 0)/\sigma_N(V = 0) \approx 1.6.$$

Thus, in the worst-case scenario, $\sigma_C > 3\sigma_N$, which indicates that contact conductance dominates junction conductance in the second multiplicative factor, and normalized conductance

$$\frac{G(V)}{G_N(V)} \approx \frac{\sigma_S(V)}{\sigma_N(V)}.$$

A comparison between the nodeless s-wave fit and the nodal-superconductor fit is depicted Fig. S9c. There are two dip features near $V_s$ = ±0.5 mV that are not captured by the BTK model, which are thought to originate from critical current effects[14]. Besides this shared discrepancy, both fits match the measured data at large bias voltages, but only the nodal fit matches near zero bias. Thus, these fits of the Andreev spectrum, considered in conjunction with fits of the tunneling spectra presented in previous sections and in the main text, indicate a nodal nature of the superconducting order parameter in this system.

**F. Temperature-dependence of the excess current in PCS data**

For a perfectly transparent ($Z = 0$) point-contact junction between a normal-metal tip and a superconducting sample, Andreev reflection results in a two-fold increase in the conductance at zero-bias, measured with respect to the normal state, due to the conversion of impinging electrons into transmitted Cooper pairs and reflected holes at the boundary. In the point-contact measurement in ballistic limit, this process enhances the current flowing through the circuit when $V > \Delta$, which is called the "excess current"[15]. According to the BTK theory, the excess current in an isotropic BCS superconductor should be proportional to $\Delta \sigma_N$, which is determined by the barrier strength parameter $Z$. The data in main text Fig. 3c shows a temperature-dependent dip in the normal conductance around $V_s = \pm 0.75$ mV, which makes it difficult to extract the excess current contribution directly from the measured sample current. In addition, we would like to neglect critical current effects described in Section E at higher bias voltages, which would also impede the extraction of the pure excess current from Andreev reflection.

In order to remove the combined conductance contribution from the normal-state and the critical current effect in the excess current, we estimate the excess current at $V_s = 2\Delta_{AR}$, as extracted from the BTK fit performed at optimal doping, base temperature, and zero magnetic field. The excess current, shown in main text Fig. 3c, makes use of an estimate of the normal-state conductance via a linear interpolation between two spectra taken when MATBG is metallic: one taken at 200 mT and 299 mK, and one taken at 0 T and T = 1.3 K (Fig. S10a). These spectra do not show signatures of Andreev reflection. By subtracting each linearly interpolated normal conductance estimate from each Andreev spectrum at a given temperature, and subsequently integrating this new spectrum from $V_s = 0$ V to $V_s = 2\Delta_{AR}$, the excess current is calculated as follows:

$$I_{exc}(V_s) = I_S(V_s) - I_N(V_s) = \int_0^{V_s} (\sigma_S(V') - \sigma_N(V'))dV'.$$

This plot shows a linear temperature-dependence of the excess current in MATBG $I_{exc} \sim \left(1 - \frac{T}{T_C}\right)$, which does not agree with the isotropic BCS expectation, where $I_{exc,BCS} \sim \sqrt{1 - \frac{T}{T_C}}$. To confirm that this linearity shown in main text Fig. 3d does not come from the normal-state conductance interpolation, we plot the $I_S(V_s = 2\Delta_{AR})$ in Fig. S10b, which also shows a linear-in-T dependence of the excess current.

In addition, we extract the energy gap $\Delta_{AR}$ using a BTK fitting analysis of the temperature-dependent point-contact spectra shown in Fig. 2d of the main text. We employ a similar background subtraction to the second method discussed above, this time dividing by the linearly interpolated normal-state conductance spectrum for each Andreev spectrum (the BTK model accepts the dimensionless input $G/G_N(V)$). We fix the barrier transparency parameter $Z = 0.1$, which was extracted from the base-temperature Andreev spectrum (Fig. S10c). Fig. 3d of the main text show $\Delta_{AR}$ extracted from this BTK fitting analysis, where error bars represent one standard deviation. We find excellent agreement between the Andreev energy gap estimate of $T_c \approx 1.1$ K and the linear excess current estimate of $T_c \approx 1.15$ K, which provides internal consistency for our interpretation.

We note that the temperature values in Fig. 2d of the main text are readings from a calibrated $RuO_2$ thermometer embedded in the STM head a few centimeters away from the sample. The $T = 1.3$ K and $T = 300$ mK PCS measurements are steady-state temperatures. However, the PCS measurements at intermediate temperatures are not steady-state, but are acquired while slowly cooling the microscope from 1.3 K to base temperature (i.e. spectra for intermediate temperatures are acquired without settling at each individual temperature). However, as soon as the thermometer value reached 300 mK, the measured PCS zero-bias conductance immediately stopped increasing in intensity (i.e. the Andreev spectrum stopped changing), implying good agreement between the sample and thermometer temperatures.

**G. Spatial dependence of tunneling spectroscopy on AA sites and AB/BA sites**

To examine the spatial dependence of the tunneling spectroscopy, we acquired STS on different AA sites within a region of our device with a uniform twist angle and strain condition. The tunneling spectra in Fig. S12 b-e do not show significant spatial dependence. The energy and the shape of the tunneling gaps are similar to each other. These are indicative of the homogeneity of our spectroscopic measurements across the moiré superlattice, and helps us correlate our measurements to those of global transport measurements. We also measured tunneling spectra on AB/BA sites in Device B (Fig. S13a and S13b). Tunneling spectroscopy on AB sites also show a tunneling gap between $-3 < \nu < -2$ that is quite similar to that observed on AA sites. Figs. S13c-e show the results of a spectroscopy line cut across an AA site (from an AB site to a BA site). The size of the $-3 < \nu < -2$ tunneling gap is independent of the spatial position.

**H. Pseudogap regime for $-3 < \nu < -2$ at high temperatures**

At millikelvin temperatures, we observe a V-shaped gap in STS that is coincident with an Andreev reflection spectrum in PCS within the superconducting density regime of MATBG. In several samples, we have identified a suppression of the DOS at the Fermi level in this density range at temperatures above the critical transition temperature of MATBG. Figs. S14a-c show tunneling d$I$/d$V$($V_s$, $V_g$) acquired at high temperatures, all above 4.1 K, that show a suppression in the valence band at $E_F$. We interpret this suppression as a signature of a high-temperature pseudogap phase.

**I. Topographic analysis of MATBG aligned to hBN**

We acquired STM topographic images (main text Fig. 5b) at various values of the sample bias and gate voltage in order to distinguish the G-G and G-hBN moiré patterns. When the MATBG flat bands are emptied ($V_g$ = -8 V), topographic images of the filled states highlight the G-hBN moiré (main text Fig. 5b; bottom left). This image shows the typical triangular lattice of darkened regions (carbon-boron sites) that are seen when monolayer graphene is aligned to hBN[16–20], while the AA sites of the G-G moiré are significantly less prominent because the spectral weight of the flat bands (which are empty) are highly localized to the AA sites. On the other hand, when the MATBG valence flat band is filled ($V_g$ = 0 V), the AA sites of the G-G moiré are highly prominent (main text Fig. 5b, top right). This image shows the typical triangular lattice of bright regions (AA sites) that is seen in twisted bilayer graphene. From the moiré length scale seen in these images, we deduce that the G-G twist angle is 1.08°, while the G-hBN twist angle is 0.5 ± 0.1° (assuming a graphene-hBN atomic lattice constant ratio of 1.017[21]).

Overall, these topographic images show that when the G-G and G-hBN moiré length scales are similar, perfect commensurability between the two moiré patterns is a physically possible scenario, in contrast to the formation of a super-superlattice documented previously[22]. Theoretical studies[21,23,24] have suggested that moiré-commensurability is important for realizing a zero-magnetic-field Chern insulating state, which we discuss in Section J.

**J. Zero-field Chern insulator in moiré-commensurate MATBG aligned to hBN**

In the main text, we noted that the insulating gap at $\nu$ = +3 has a Chern number C = 1. We provide evidence for this claim here through a method presented in Refs.[25,26]. Figs. S15a,d show two d$I$/d$V$($V_s$, $V_g$) data sets (acquired on AAb sites) from two similar regions of Device C that are both moiré-commensurate, and Figs. S15b,c are zoomed-in d$I$/d$V$($V_s$, $V_g$) near $\nu$ = +3. Fig. S15e shows the $\nu$ = +3 gap moving to higher gate voltages as the out-of-plane magnetic field is increased from 0 T to 4 T. Since $\frac{dn}{dB} = \frac{\sigma_{xy}}{e} = \frac{C}{\Phi_0}$ (where $\Phi_0$ is the magnetic flux quantum) for an insulating phase[27], we conclude the $\nu$ = +3 gap corresponds to a C = +1 insulating phase at zero magnetic field, in agreement with a previous transport study[28]. In addition, we also observe an insulating phase with C = -2 of unknown origin.

## K. Summary of the superconducting states in six devices

A summary of the experimental data of the superconducting states shown in this study are listed here:

| Device | Angle (°) | $V_g$ (V) | $T_c$ (K) | $B_c$ (mT) | $2\Delta_T$ (meV) | $2\Delta_{AR}$ (meV) |
|---|---|---|---|---|---|---|
| A | 1.13 | -25.8 | | | 1.8 | |
| A' | 1.01 | -22.8 | 1.2 | 50 | 2.8 ($2\Delta_T/k_BT_c = 27$) | 0.6 ($2\Delta_{AR}/k_BT_c = 5.8$) |
| B | 1.06 | -25.6 | | | 2.9 | |
| D | 1.06 | -30.5 | | 200 | | 1.4 |
| F | 1.1 | -25.6 | | | 1.3 | |
| G | 1.17 | -30 | 1.2 | 75 | | 0.6 |

## L. Tunneling and PCS parameters for main text figures

The tunneling setpoint and PCS parameters for the main text figures are listed here:

| Figure | $V_s$ (mV) | I (nA) | $V_g$ (V) | $V_{rms}$ (mV) | $f_{rms}$ (Hz) | T (mK) | B (T) |
|---|---|---|---|---|---|---|---|
| 1b | -70 | 0.3 | -25.8 | | | | |
| 1c | -80 | 0.8 | | 0.5 | 1262.7 | 250 | 0 |
| 1d | -80 | 1.8 | | 0.15 | 1262.7 | 200 | 0 |
| 1e (upper) | -80 | 1.8 | | 0.15 | 1262.7 | 200 | 0 |
| 1e (lower) | -80 | 1.8 | | 0.05 | 1262.7 | 200 | 0 |
| 1f | -80 | 1.5 | | 0.2 | 4121 | 200 | 0 |
| 1g (both) | -80 | 1.5 | | 0.06 | 4121 | 200 | 0 |
| 2b, 2c, 2e, 2g-PCS | | | | 0.05 | 381.7 | 270 | |
| 2d, 2f | | | | 0.05 | 381.7 | | 0 |
| 2g-STS | -80 | 0.5 | | 0.5 | 1262.7 | 280 | 0 |
| 3e | -100 | 0.5 | | 0.5 | 381.7 | 4.1 | 0 |
| 4a | -80 | 1.8 | | 0.15 | 1262.7 | 200 | 0 |
| | -70 | 1.7 | | 0.15 | 1262.7 | 185 | 0.5 |
| | -80 | 1.5 | | 0.15 | 1262.7 | 250 | 1 |
| 4b | -80 | 1.5 | | 0.2 | 4121 | 200 | 0 |
| | -80 | 1.5 | | 0.2 | 4121 | 200 | 3 |
| | -80 | 1.0 | | 0.2 | 4121 | 200 | 6 |
| 5a | -400 | 0.1 | 8 | | | | |
| 5b | -300 | 0.01 | 0 | | | | |
| | -100 | 0.01 | 0 | | | | |
| | -300 | 0.01 | -8 | | | | |
| | -100 | 0.01 | -8 | | | | |
| 5c-STS | -80 | 1.5 | | 1 | 381.7 | 190 | 0 |
| 5c-PCS | | | | 0.2 | 381.7 | 255 | 0 |

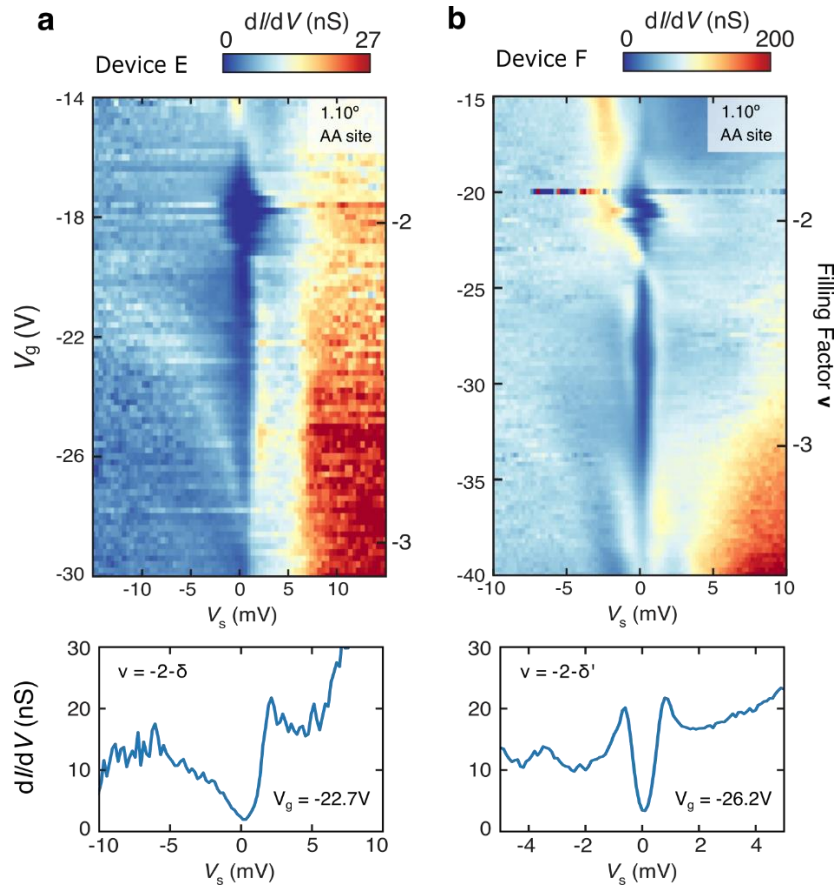

**Figure S1 | Two additional observations of the tunneling gaps near $\nu = -2$ and between -3 < $\nu$ < -2. a,** Tunneling d$I$/d$V$($V_s$, $V_g$) and spectroscopy line cut at $V_g$ = -22.7 V acquired at the center of an AA site in Device E (angle : 1.10°) at zero magnetic field. An insulating gap with Coulomb charging effects opens near $V_g$ = -15 V and closes near $V_g$ = -19 V (i.e. surrounding $\nu$ = -2). A second gap reopens below $\nu$ = -2 and persists to near $\nu$ = -3. The gap between -3 < $\nu$ < -2 is less clear due to the slightly compromised tip condition. **b,** Similar tunneling d$I$/d$V$($V_s$, $V_g$) and spectroscopy line cut at $V_g$ = -26.2 V acquired at the center of an AA site in Device F (angle : 1.10°, strain : 0.1%) at zero magnetic field. A gap with Coulomb charging effects opens near $V_g$ = -18 V and closes near $V_g$ = -23 V (i.e. surrounding $\nu$ = -2). A second gap reopens below $\nu$ = -2 and persists to near $\nu$ = -3. Between -3 < $\nu$ < -2, spectra show prominent coherence peaks that flank both sides of the gap. In both spectroscopy line cuts in **a** and **b**, the tunneling gaps do not show conventional s-wave features.

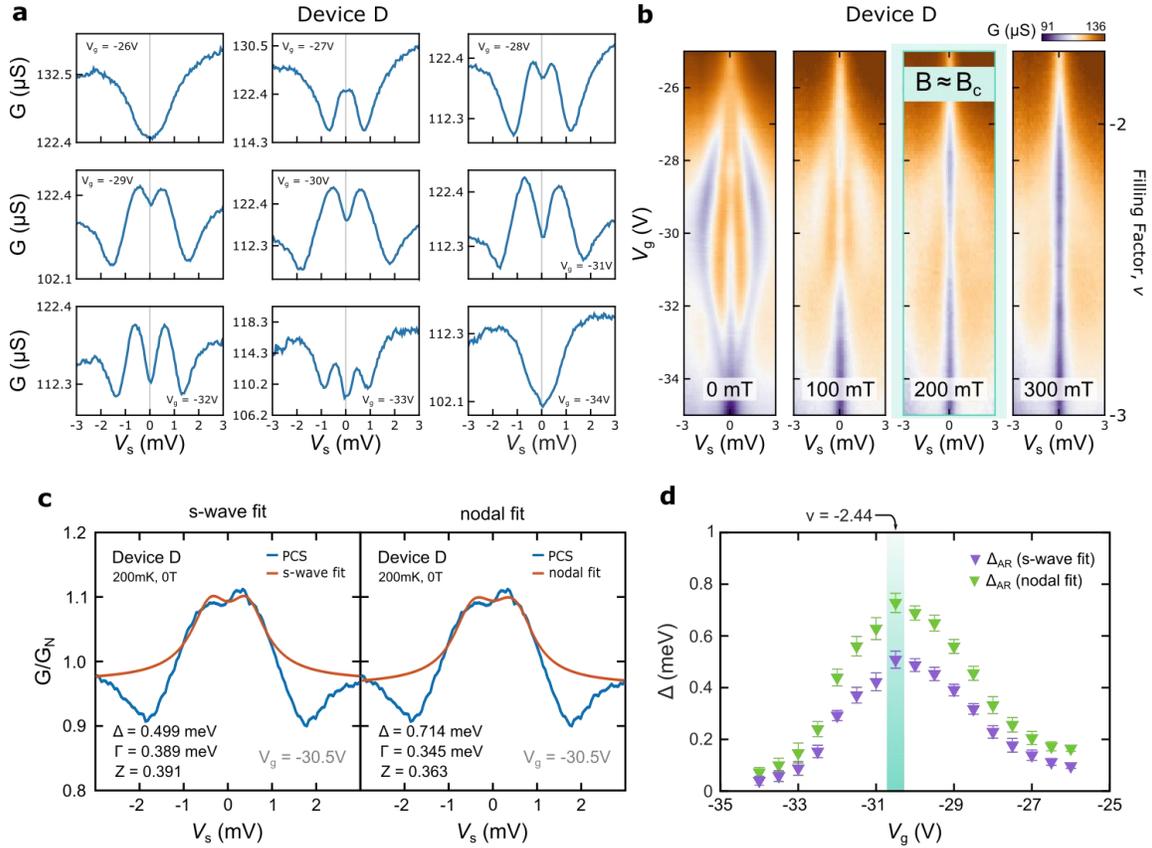

**Figure S2 | Large Andreev energy gap in point-contact spectroscopy in Device D. a,** Point-contact spectra G($V_s$) in Device D (angle : 1.06°) at $T$ = 200 mK, $B$ = 0 T, and at nine gate voltages between $V_g$ = -26 V and $V_g$ = -34 V. In this device, the Andreev spectrum onsets smoothly, widening in energy to an optimal doping near $V_g$ = -30.5 V, and then closes. **b,** Point-contact spectroscopy G($V_s, V_g$) in Device D at four magnetic-field strengths between 0 T and 300 mT. The critical magnetic field, as measured by the Andreev spectrum, is slightly higher than 200 mT. A persistent zero-bias suppression at 300 mT may indicate residual resistance in either the tip-sample junction or the sample contact electrode. **c,** BTK fitting analysis of the normalized zero-field Andreev spectrum at optimal doping, fit with s-wave and nodal gap symmetries. **d,** Extracted Andreev energy gaps $\Delta_{AR}$ as a function of gate voltage $V_g$ from s-wave and nodal fits, which show an expected dome shape. The gap in this device reaches an optimal value of $\Delta_{AR,s-wave}$ = 0.50 meV and $\Delta_{AR,nodal}$ = 0.71 meV at filling $\nu$ = -2.44.

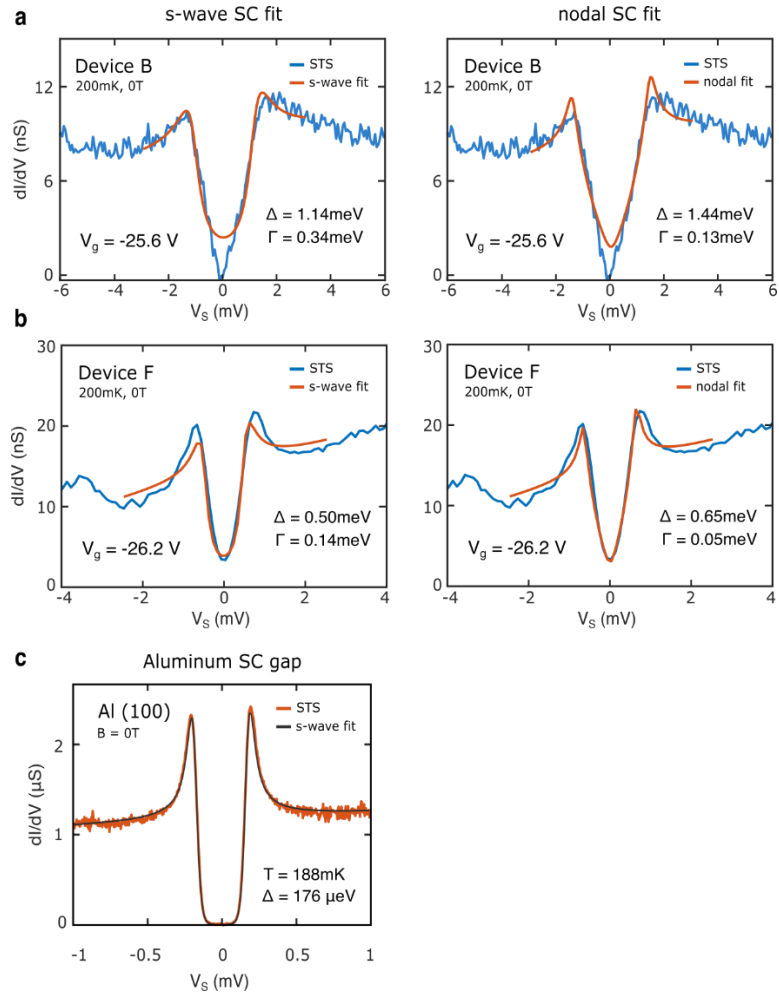

**Figure S3 | Dynes-function fitting analysis of the tunneling gap between -3 < $\nu$ < -2. a,** s-wave-superconductor fit (left) and nodal-superconductor fit (right) of the tunneling spectrum d$I$/d$V$($V_s$) at $V_g$ = -25.6 V in Device B using the Dynes function. The energy gap $\Delta$ and the quasiparticle lifetime broadening $\Gamma$ are free parameters. **b,** same as **a**, but at $V_g$ = -26.2 V in Device F. In both cases, nodal-SC fits capture spectroscopic features better than the s-wave fits. **c,** BCS s-wave fit of tunneling data on a superconducting Al(100) surface at base temperature using the Dynes function. $\Delta$, $\Gamma$, and T are all free fitting parameters. Aluminum has a superconducting transition temperature similar to MATBG. The broadening observed in this tunneling spectra was used as a calibration of the energy resolution (~50 µeV) of our homebuilt dilution-refrigerator STM.

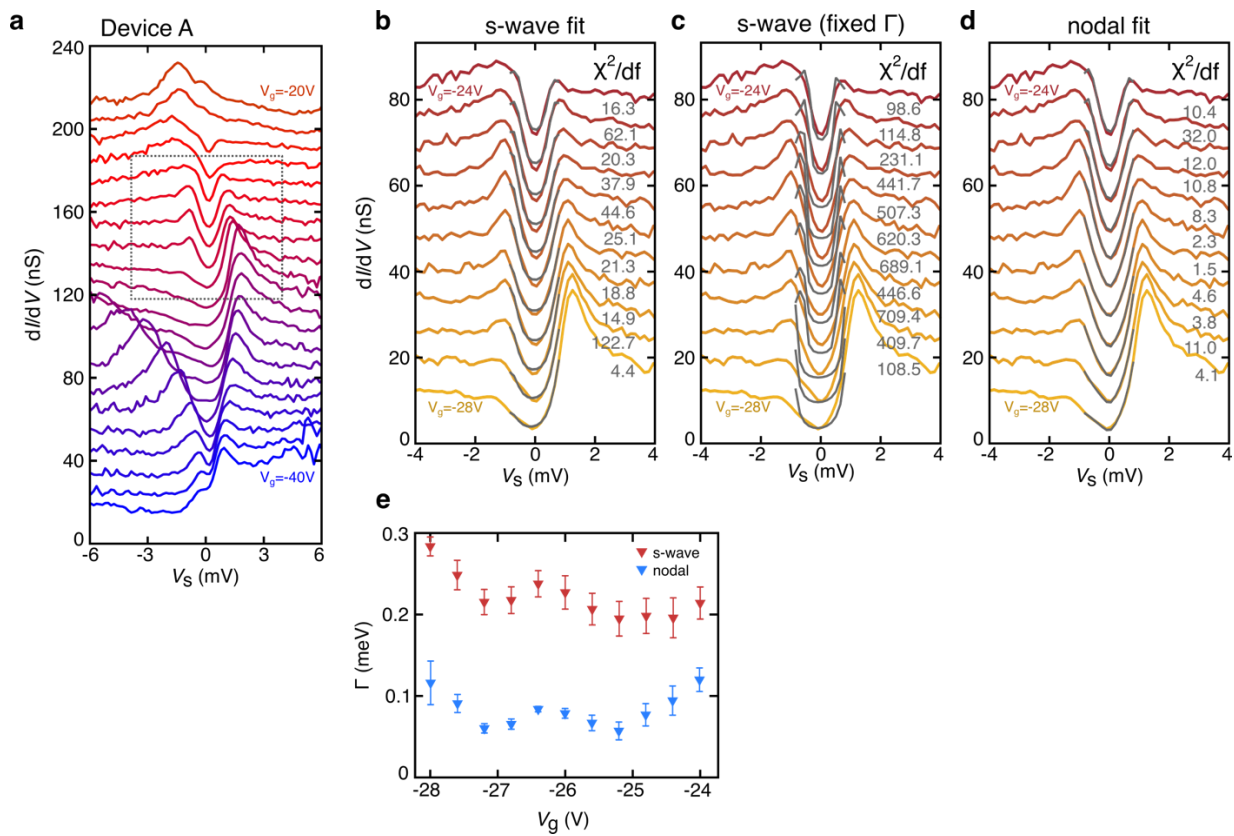

**Figure S4 | Dynes-function fitting analysis of the low-energy parts of tunneling spectra at various gate voltages throughout the -3 < $\nu$ < -2 gap. a,** Tunneling d$I$/d$V$($V_s$) spectra on an AA site in Device A at 200 mK and zero magnetic field, shown for gate voltages between $V_g$ = -20 V and $V_g$ = -40 V. Curves are offset by 10 nS for clarity. The dashed-line box highlights a set of gaps between -3 < $\nu$ < -2. **b,** s-wave superconductor fits and **c,** nodal-superconductor fits of the low-energy part of eleven tunneling spectra taken from **a** from $V_g$ = -24 V to $V_g$ = -28 V. Curves are offset by 6.7 nS for clarity. For evaluating the goodness of each fit, $\chi^2$ divided by the number of degrees of freedom ($\chi^2/df$) is evaluated for each curve. When comparing s-wave and nodal fits performed on the same spectrum at every gate voltage, the nodal fits are consistently superior, with each nodal fit showing a lower $\chi^2/df$ value than the corresponding s-wave fit. **d,** same as **b**, but with quasiparticle lifetime broadening $\Gamma$ of nodal fit for corresponding $V_g$. **e,** quasiparticle lifetime broadening factors $\Gamma$ estimated from s-wave fits and nodal fits.

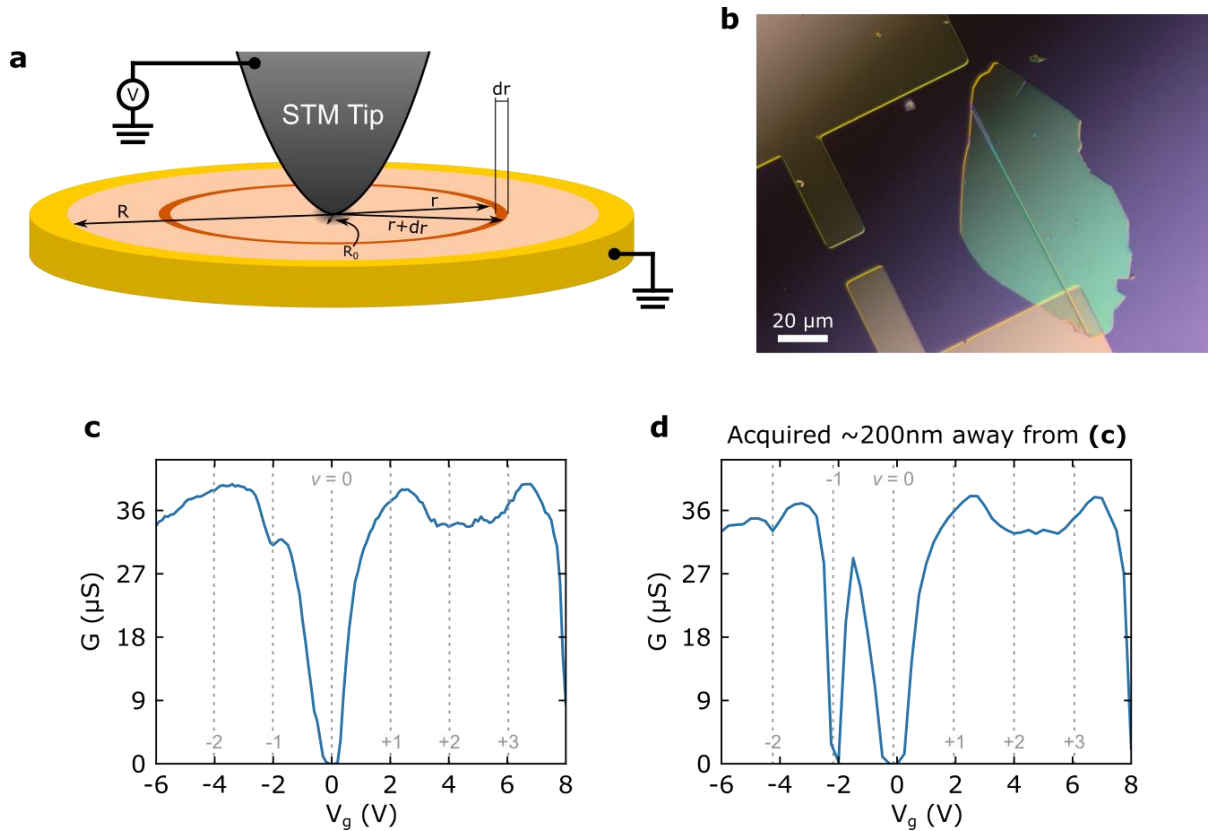

**Figure S5 | The local nature of density-tuned point-contact spectroscopy (DT-PCS). a,** Schematic diagram of the simple intuition model for the locality of our PCS method. A voltage $V$ is placed on an STM tip of contact radius $R_0$, which touches the center of a circular sample of radius $R$, whose outer perimeter is held at ground ($V = 0$ V). **b,** Nomarski differential interference contrast image of Device A. **c,** Point-contact G($V_g$) at $V_s$ = 0 V, taken in aligned Device C. A very weak suppression of the conductance is observed at $\nu$ = -1 at this location. **d,** Point-contact G($V_g$) at $V_s$ = 0 V, taken in aligned Device C. A strong, nearly complete suppression of the conductance is observed at $\nu$ = -1 at this location, roughly 200 nm away from the data location in **c**.

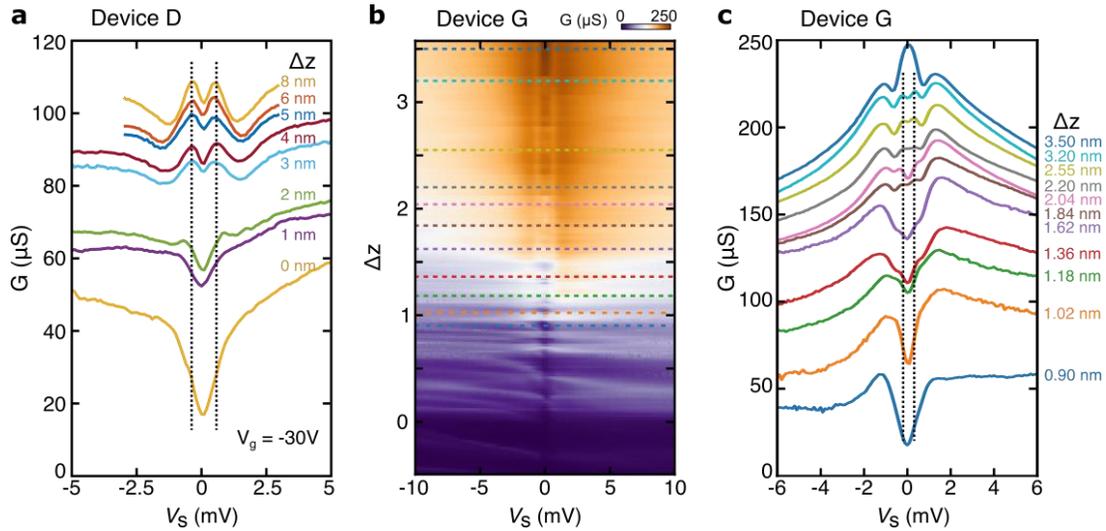

**Figure S6 | Point-contact measurement vs. tip depth. a,** Point-contact spectra G($V_s$) obtained at $V_g$ = -30 V, $B$ = 0 T, and T = 200 mK in Device D. Between each spectrum, the STM was vertically displaced by 1-2 nm. Here, $\Delta z$ is the tip height relative to the bottom curve (i.e., the largest tip-sample separation shown). As the STM tip comes into contact with the superconducting MATBG sample, the V-shaped dip transforms into a double-peaked Andreev spectrum. The barrier strength parameter Z for this tip contact is larger than that of main text Fig. 2 but similar to that of Fig. S2. The energy difference of the two peaks of the Andreev spectral feature which is around ~$2\Delta_c$ and is independent of the vertical tip displacement over $\Delta z$ ~ 5 nm (spectra labeled "3 nm" to "8 nm"). **b,** Point-contact spectra G($V_s$) obtained at $V_g$ = -30.4 V, $B$ = 0 T, and T = 265 mK in Device G. Larger values of $\Delta z$ correspond to the tip displaced deeper into the surface. The evolution of the PCS with $\Delta z$ is nonmonotonic, possibly because the tip deforms as it is pressed into the surface. **c,** PCS line cuts at select values of $\Delta z$ indicated by the dashed lines in **b**.

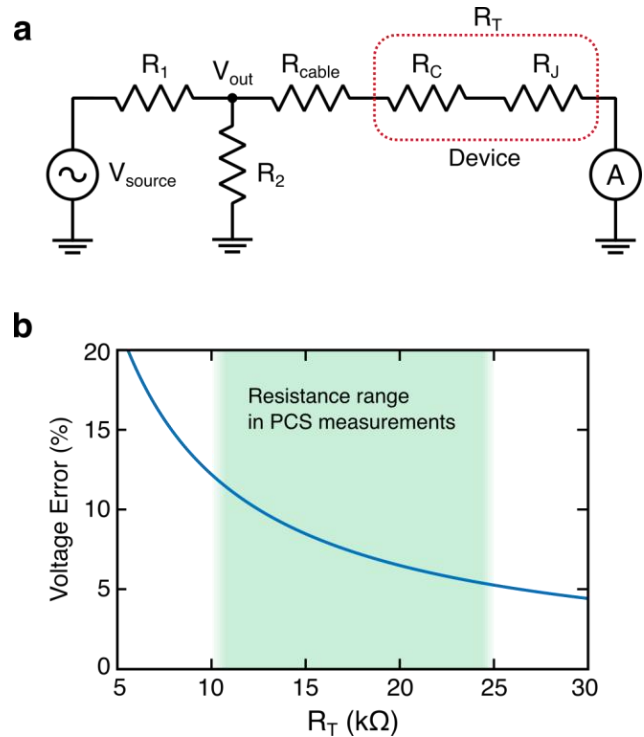

**Figure S7 | Circuit diagram for the point-contact measurement. a,** Circuit diagram for the point-contact measurement. $R_1 = 100$ kΩ and $R_2 = 1$ kΩ for the 1:100 voltage divider for the DC source voltage $V_{source}$, $R_{cable} = 400$ Ω is the cryostat cable resistance, and $R_C$ and $R_J$ are the electrode-device contact resistance and the tip-sample junction resistance, respectively. We refer to the series combination of $R_C$ and $R_J$ as $R_T$. In PCS, $R_T$ is comparable to $R_2$, so the output voltage of the voltage divider $V_{out}$ can be smaller than 1/100 of the source voltage. **b,** The voltage error with respect to the expected sample bias voltage $V_s = V_{source}/100$, plotted as a function of $R_T$. In our measurements, $R_T$ falls within the range between 10 kΩ and 25 kΩ, which may cause overestimation of Andreev energy gaps by ~10%.

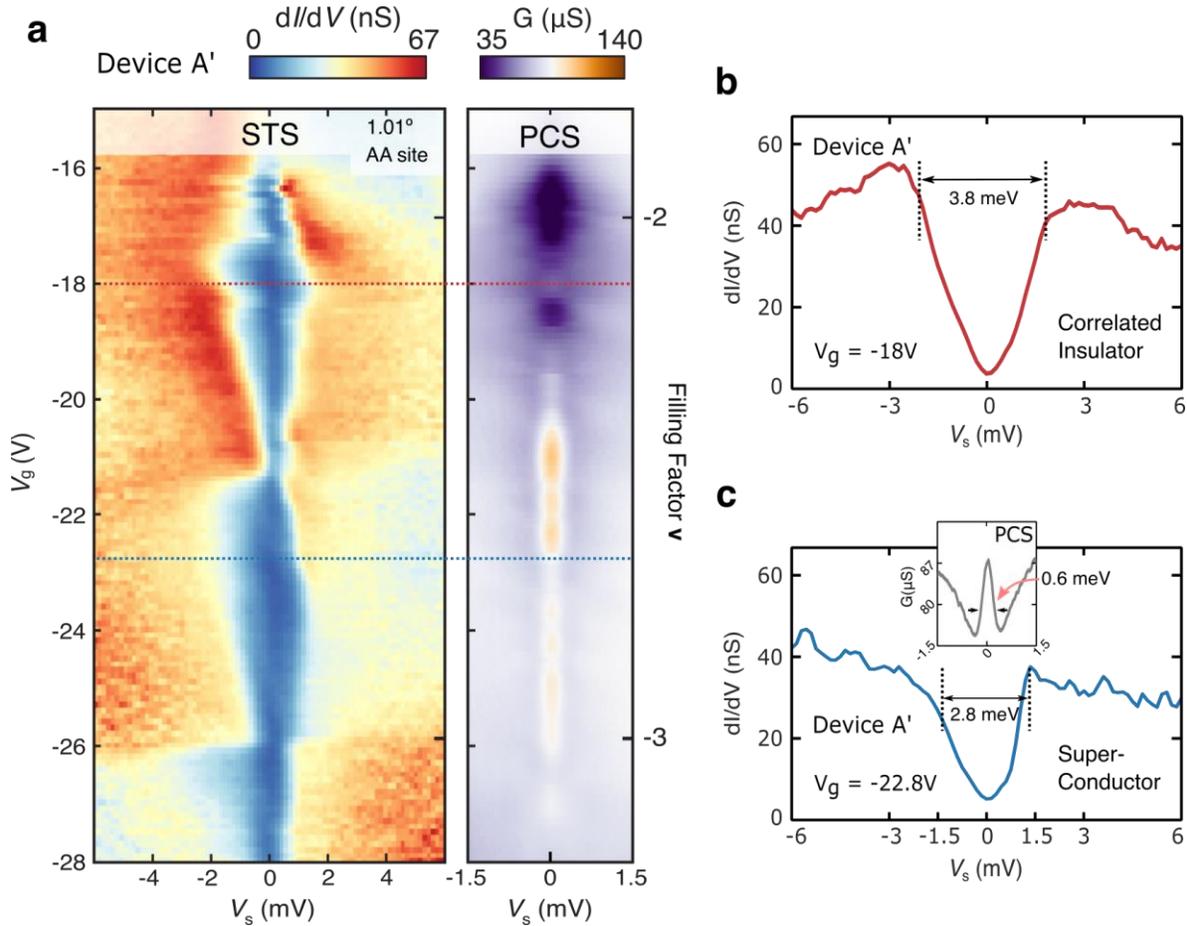

**Figure S8 | Tunneling spectroscopy and point-contact spectroscopy at -3 < $\nu$ < -2 in Device A'. a,** Side-by-side tunneling d$I$/d$V$($V_s$, $V_g$) into an AA site and point-contact G($V_s$, $V_g$) in the same location in Device A'. d$I$/d$V$($V_s$, $V_g$) shows the opening and closing of the correlated insulating and superconducting phases at the Fermi level. G($V_s$, $V_g$) allows us to differentiate between the origins of these two tunneling gap region, where Andreev reflection (orange peak near $E_F$) occurs only in the superconducting phase, in contrast to the gapped spectroscopic feature (purple dip near $E_F$) that marks the onset of the correlated insulating phase. **b,** Tunneling spectroscopy of the correlated insulator gap at $V_g$= -18 V. **c,** Tunneling spectroscopy of the superconducting gap at $V_g$= -22.8 V. Point-contact spectrum (inset in **c**) of the corresponding tunneling gap of the superconducting phase shows a clear difference between the two energy scales $\Delta_{AR}$ and $\Delta_T$.

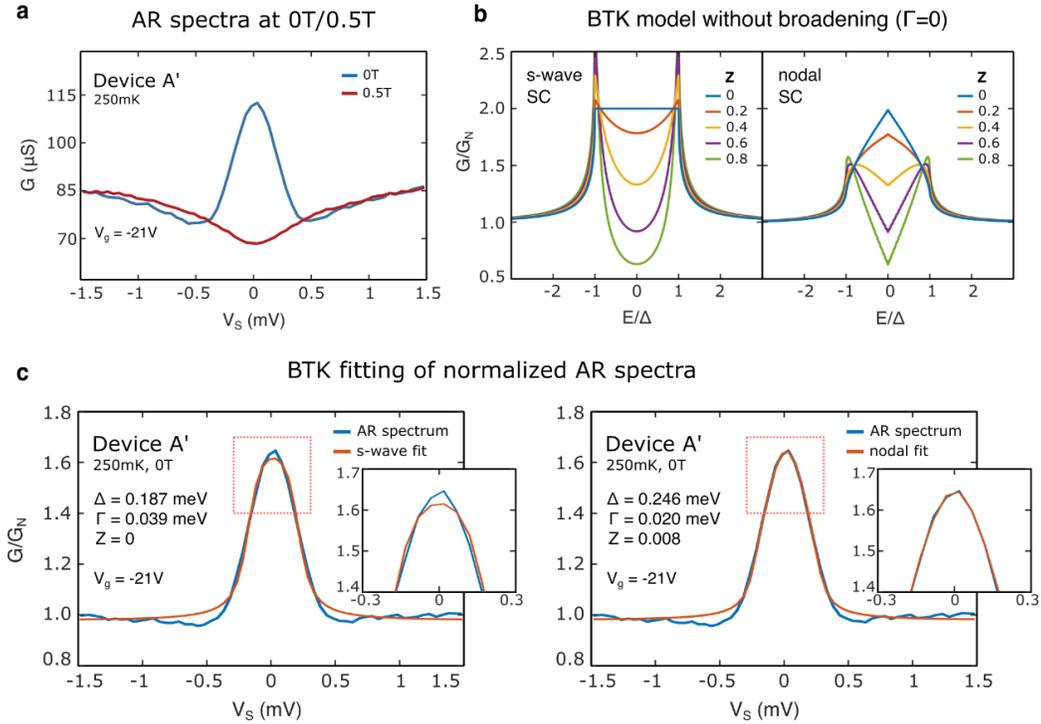

**Figure S9 | Blonder-Tinkham-Klapwijk (BTK) fitting analysis of the Andreev reflection spectrum at -3 < $v$ < -2. a,** Point-contact spectrum G($V_s$) at $V_g$ = -21 V at $B$ = 0 T (blue) and $B$ = 0.5 T (red) in Device A'. The blue curve is peaked at zero-bias, indicative of the Andreev reflection process at $B$ = 0 T. The red curve shows a weak minimum at zero-bias, indicative of residual resistance in the tip-sample junction or the sample contact, as the sample has entered the normal state at $B$ = 0.5 T. Normalized data $G/G_N$ for the BTK model is obtained by dividing the 0 T superconducting-state data by the 0.5 T normal-state data. **b,** Comparison of BTK simulations of c-axis transport between a normal-metal tip and an s-wave (left) or nodal (right) superconducting sample, plotted for various barrier strength parameters $Z$. The s-wave model shows a flat-top-shaped spectrum when $E < \Delta$ with a doubled conductance at $Z = 0$, while the nodal-superconductor (nodal-SC) model shows a point-shaped spectrum with a doubled conductance only at $E = 0$ and $Z = 0$. **c,** BTK-model fitting of a normalized spectrum from the data in **a,** using an s-wave (left) and a nodal (right) superconducting order parameter. $\Delta$, $\Gamma$, and $Z$ are free fitting parameters. Red dashed-line boxes highlight the apex of the peaks in each plot, where there is the largest discrepancy between the s-wave and the nodal fits (insets). In addition, the broadening in the s-wave scenario is large ($\Gamma > 0.2\Delta$), while the broadening in the nodal scenario is reasonable ($\Gamma \approx 0.08\Delta$).

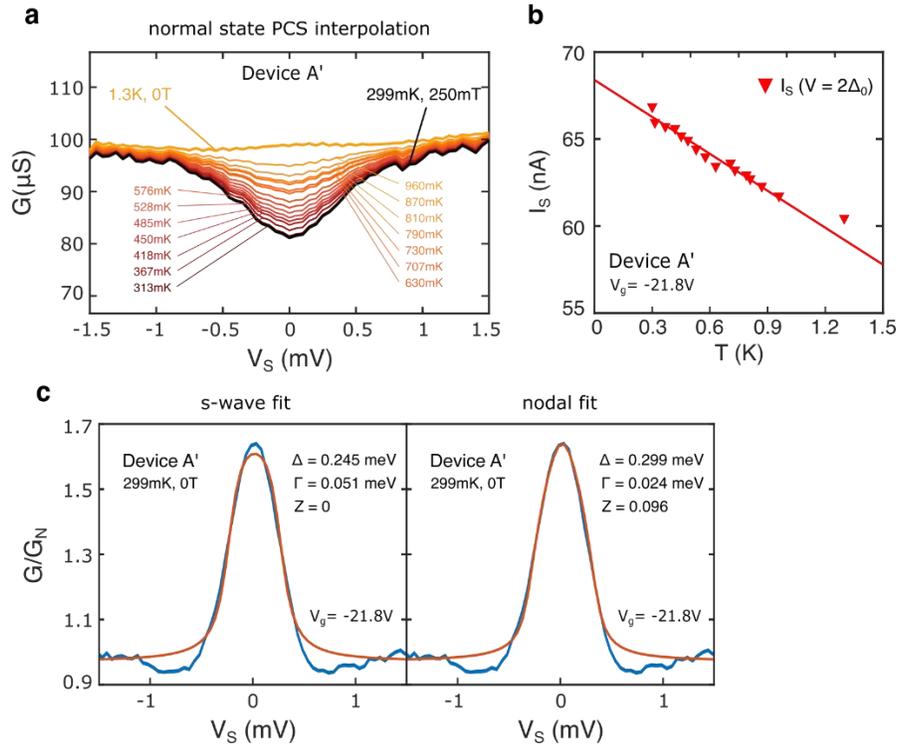

**Figure S10 | Temperature-dependent (BTK) fitting analysis of the Andreev reflection spectrum at -3 < $\nu$ < -2. a,** Normal state point-contact spectra G($V_s$) estimated via linear interpolation between a spectrum taken in Device A' at T = 299 mK, B = 250 mT and a spectrum taken at T = 1.3 K, B = 0 T. The excess current depicted in main text Fig. 3d is calculated by subtracting the integral of each Andreev spectrum with respect to the bias voltage from the integral of these interpolated normal-state spectra. **b,** Temperature dependence of the superconducting-state current $I_s$ at V = $2\Delta_0$, where $\Delta_0$ = 0.3 meV is the gap size extracted from the PCS data at 299 mK. This shows a linear relationship between $I_s$ and T. **c,** BTK fits of the normalized PCS spectra at T = 299 mK, used to estimate the barrier strength parameter $Z$ in s-wave (left) and nodal (right) superconductor scenarios, with free fitting parameters $\Delta$, $\Gamma$ and $Z$. Nodal superconductor fits better than s-wave. Thus, we use nodal superconductor fits with Z ~ 0.1 extracted from this fit for temperature dependent BTK analysis in main text Fig. 3c.

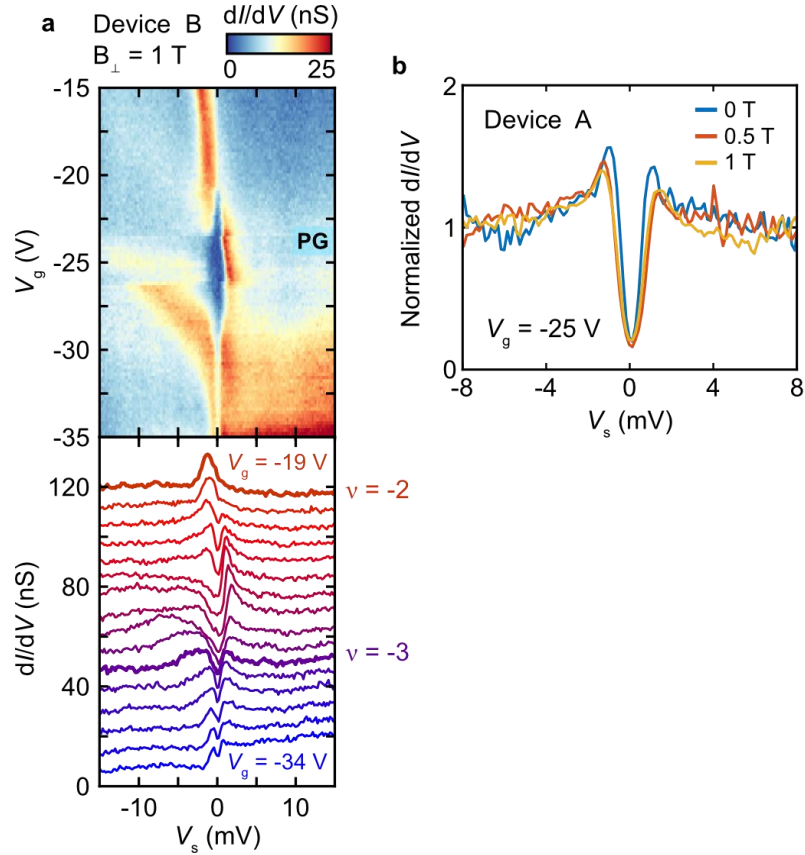

**Figure S11 | Tunneling spectroscopy at 1 T in Device B and comparison of tunneling spectra at various magnetic fields. a,** Tunneling $dI/dV(V_s, V_g)$ and spectroscopy line cuts acquired at the center of an AA site in Device B for $V_g$ = -19 V to $V_g$ = -34 V and for $B_\perp$ = 1 T. Curves are offset by 7.5 nS for clarity. Although this is above the critical magnetic field for superconductivity in MATBG, we observe a suppression of the zero-bias conductance, indicating the presence of the high-field pseudogap state discussed in the main text between -3 < $\nu$ < -2. Initial tunneling parameters: $V_s$ = -80 mV, $I$ = 1500 pA, $V_{rms}$ = 0.2 mV at 4.121 kHz, $T$ = 200 mK. **b,** Direct comparison of tunneling spectra $dI/dV(V_s)$, normalized for their linear background, observed in Device A at $V_g$ = -25 V. Only the spectrum observed at $B_\perp$ = 0 T was acquired in the superconducting state.

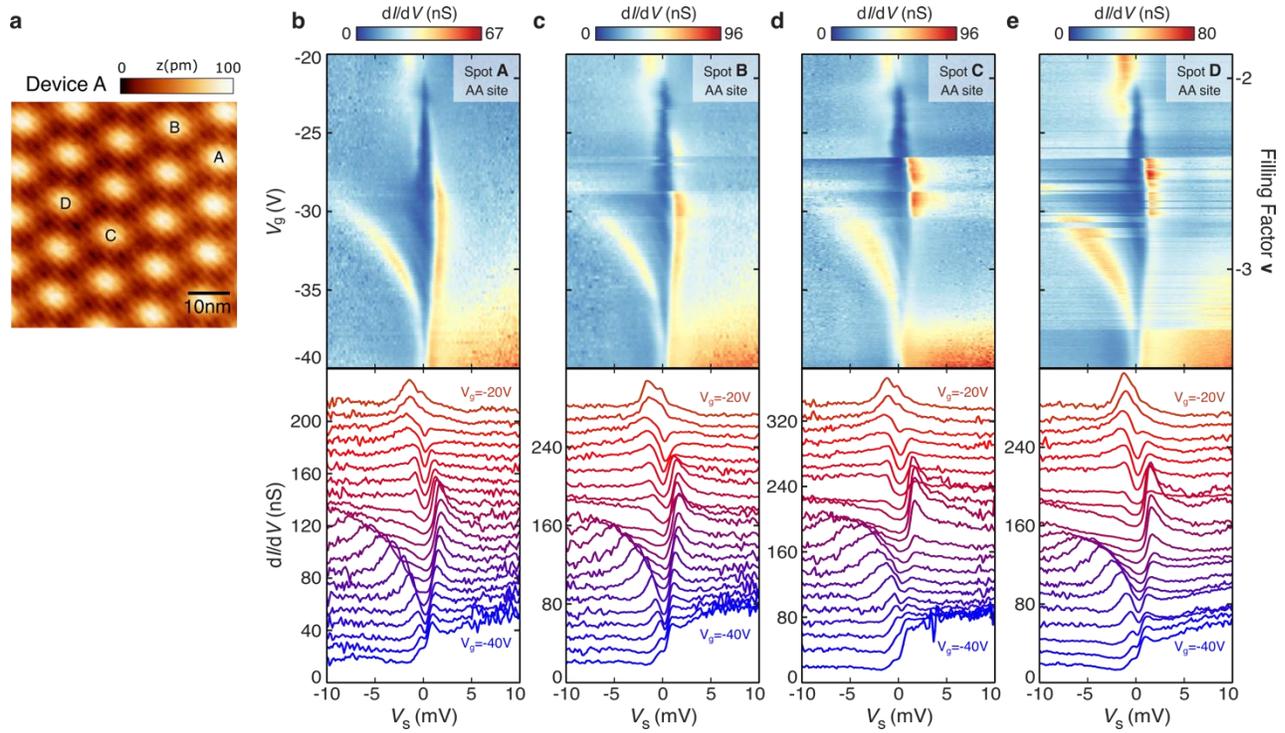

**Figure S12 | Tunneling spectroscopy at the center of different AA sites. a,** Topographic image of Device A. **b,** Tunneling $dI/dV(V_s, V_g)$ and $dI/dV(V_s)$ specra from an AA site, obtained at $B = 0$ T in Device A for $V_g = -20$ V to $V_g = -40$ V, obtained at the spot labeled "A" in **a**. **c-e,** same as **b**, but for spot B (**c**), spot C (**d**), and spot D (**e**) obtained at $B_\perp = 0.5$ T. In **b-e**, curves are offset for clarity by 10 nS, 12.8 nS, 16 nS and 12.8 nS, respectively.

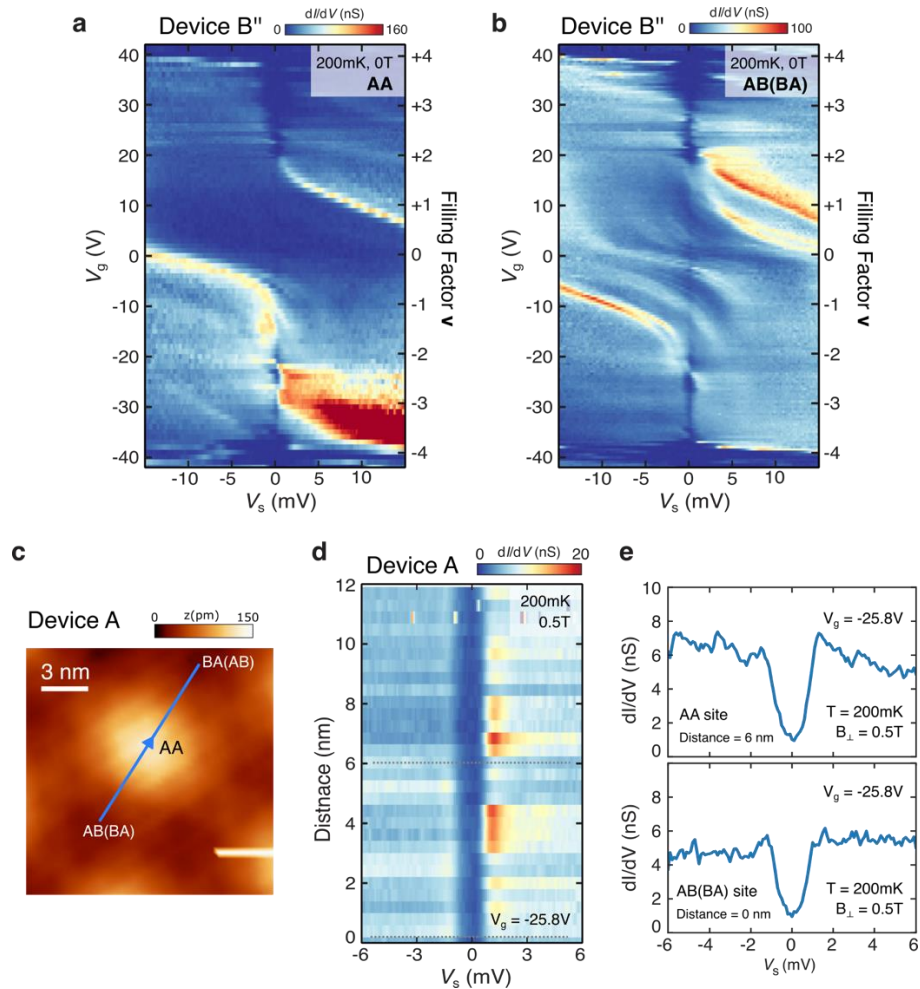

**Figure S13 | Tunneling spectroscopy at AA and AB(BA) site, and spatial dependence of the -3 < $\nu$ < -2 tunneling gap. a,** Tunneling dI/dV($V_s$, $V_g$) obtained from an AA site in Device B" (angle : 1.07°). **b,** same as **a,** but from an AB (or BA) site. Both show tunneling gaps at -3 < $\nu$ < -2. **c,** Topographic image near AA site in Device A. **d,** Line cut of dI/dV($V_s$) along the blue line drawn in **c** at $V_g$ = -25.8 V at $B_\perp$ = 0.5 T. Tunneling gap sizes are almost independent on the location. Gray dashed lines highlight the location of the AB(BA) site (distance = 0 nm) and the AA site (distance = 6 nm). **e,** dI/dV($V_s$) spectra at AA site (top), and dI/dV($V_s$) spectra at AB(BA) site (bottom).

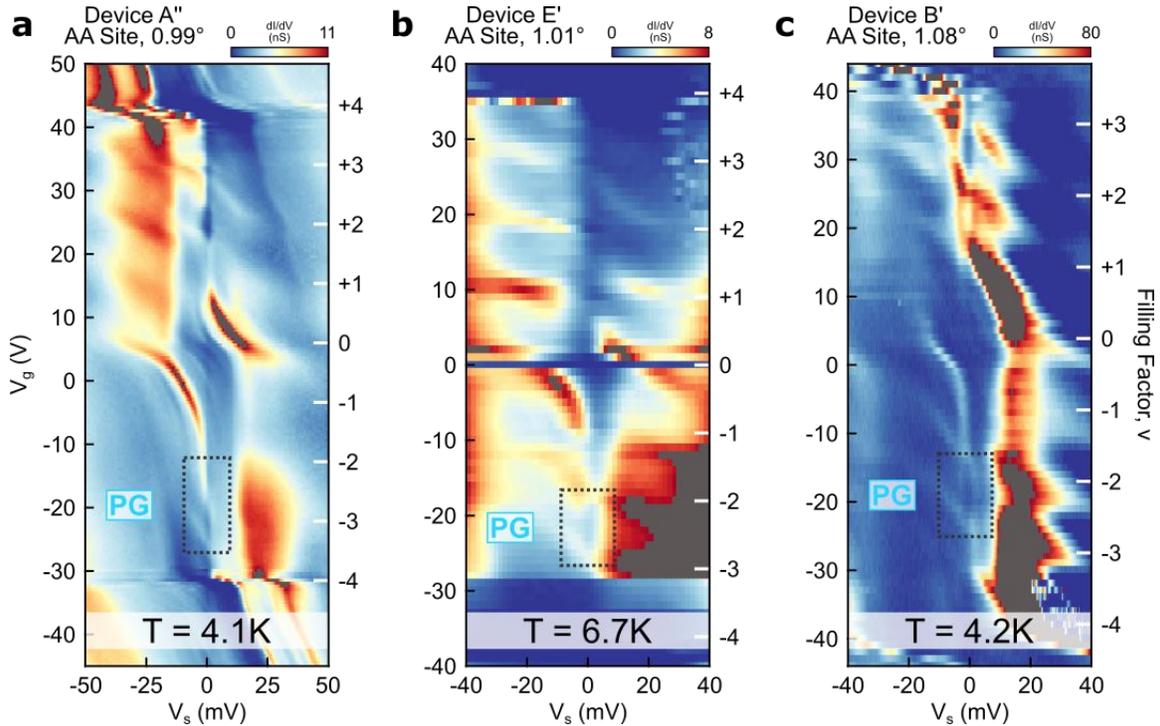

**Figure S14 | Tunneling spectroscopy of the pseudogap phase at temperatures above 4 K. a,** Tunneling $dI/dV(V_s, V_g)$ acquired at the center of an AA site in Device A'' (0.99°) at zero magnetic field and $T = 4.1$ K. A spectral suppression near $E_F$ appears in the valence flat band around $\nu < -2$, which we attribute to a high-temperature pseudogap state at this temperature. **b,** Tunneling $dI/dV(V_s, V_g)$ acquired at the center of an AA site in Device E' (very nearly the same area as Device E millikelvin data; 1.01°) at zero magnetic field and $T = 6.7$ K. A spectral suppression near $E_F$ appears in the valence flat band around $-3 < \nu < -2$, which we attribute to the pseudogap state. **c,** Tunneling $dI/dV(V_s, V_g)$ acquired at the center of an AA site in Device B' (different area in Device B; 1.08°) at zero magnetic field and $T = 4.2$ K. A gap-like spectral suppression near $E_F$ appears in the valence flat band around $-3 < \nu \leq -2$, which we attribute to the pseudogap state.

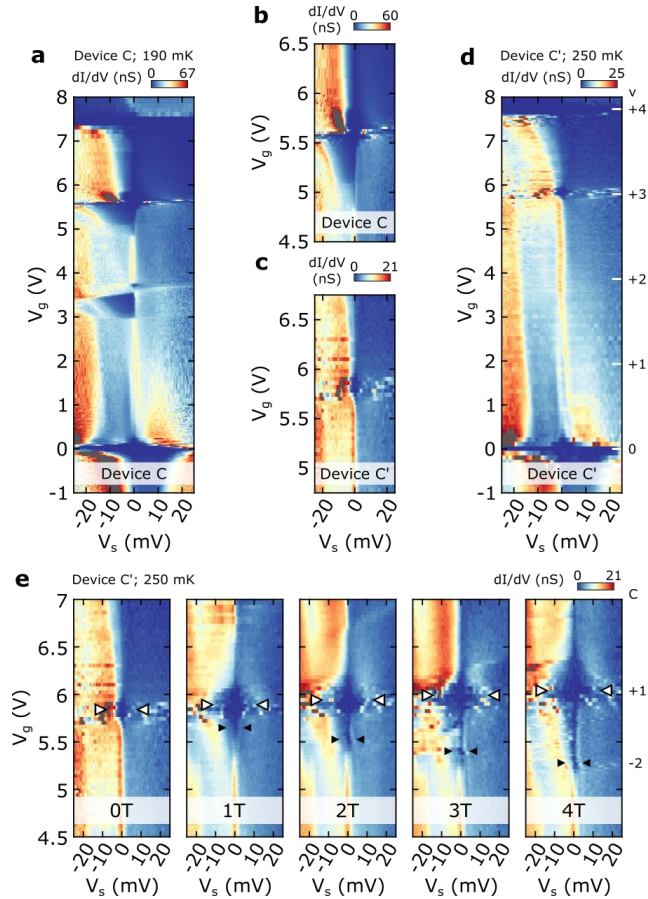

**Figure S15 | Tunneling spectroscopy and Chern characterization of moiré-commensurate MATBG aligned to hBN. a,** Tunneling $dI/dV(V_s, V_g)$ acquired at the center of an AAb site in a moiré-commensurate region of Device C (1.08° G-G twist angle, 0.1% G-G interlayer strain, 0.5 ± 0.1° G-hBN twist angle). A $C_2$-symmetry-broken gap, a correlated insulator gap, and a correlated Chern C = +1 gap appear near $\nu$ = 0, +2, and +3, respectively. **b,** Zoomed-in $dI/dV(V_s, V_g)$ from **a** to highlight the observed zero-field Chern C = +1 gap. **c,** Zoomed in $dI/dV(V_s, V_g)$ from **d** to highlight the observed zero-field Chern C = +1 gap. **d,** Tunneling $dI/dV(V_s, V_g)$ acquired at the center of an AAb site in Device C' (same as Device C, at a different sample location; 1.10°). A $C_2$-symmetry-broken gap and a correlated Chern C = +1 gap appear near $\nu$ = 0 and +3, respectively. **e,** Tunneling $dI/dV(V_s, V_g)$ at five different magnetic-field strengths between 0 T and 4 T, at around $\nu$ = +3. We observe a correlated Chern C = +1 state at and emanating from $\nu$ = +3. We also observe weaker signatures of a topological gapped phase with Chern number -2, which is first resolved at 1 T.

# SUPPLEMENTARY FIGURE S16

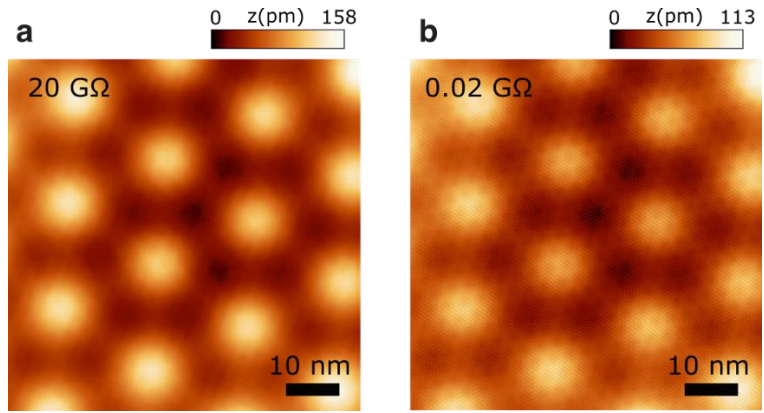

**Figure S16 | Topographic images at different tunneling setpoints. a**, Topographic image of Device G' (1.18°) at $I$ = 10 pA, $V_s$ = -200 mV (20 GΩ tunnel resistance). **b**, Same as **a** except at $I$ = 500 pA, $V_s$ = -10 mV (0.02 GΩ tunnel resistance).